\documentclass[%
 reprint,
 superscriptaddress,
 amsmath,amssymb,
 aps,
 pre,
]{revtex4-1}
\usepackage{graphicx}
\usepackage{dcolumn}
\usepackage{bm}

\let\oldhat\hat
\renewcommand{\vec}[1]{\mathbf{#1}}
\renewcommand{\hat}[1]{\oldhat{\mathbf{#1}}}

\begin{document}

\title{Generalized Archimedes' principle in active fluids}
\author{Nitzan Razin}
 \affiliation{Department of Chemical Physics, Weizmann Institute of Science, Rehovot 76100, Israel}
\author{Raphael Voituriez}%
\affiliation{Laboratoire Jean Perrin and Laboratoire de Physique Th\'eorique de la Mati\`ere Condens\'ee, CNRS/Universit\'e Pierre et Marie Curie, 75005 Paris, France}
\author{Jens Elgeti}
\affiliation{Theoretical Soft Matter and Biophysics, Institute of Complex Systems and Institute for Advanced Simulation, Forschungszentrum J\"ulich, D-52425 J\"ulich, Germany}

\author{Nir S. Gov}
\affiliation{Department of Chemical Physics, Weizmann Institute of Science, Rehovot 76100, Israel}%

\date{\today}

\begin{abstract}

We show how a gradient in the motility properties of non-interacting point-like active particles can cause a pressure gradient that pushes a large inert object. We calculate the force on an object inside a system of active particles with position dependent motion parameters, in one and two dimensions, and show that a modified Archimedes' principle is satisfied. We characterize the system, both in terms of the model parameters and in terms of experimentally measurable quantities: the spatial profiles of the density, velocity and pressure. This theoretical analysis is motivated by recent experiments, which showed that the nucleus of a mouse oocyte (immature egg cell) moves from the cortex to the center due to a gradient of activity of vesicles propelled by molecular motors; it more generally applies to artificial systems of controlled localized activity.
\end{abstract}

\maketitle


\emph{Introduction ---} The pressure applied by active particles to surfaces and objects has been a recent subject of interest \cite{Solon2015,Solon2015PRL,Nikola2016,Marchetti201634,Takatori2014,Yang2014,Yan2015,Ezhilan2015,Fily2014,Winkler2015,Mallory2014,Junot2017}. Generally this pressure depends on the details of the interaction with the wall \cite{Solon2015}. Interestingly, it has been shown that under certain conditions equilibrium-like properties of the pressure are retrieved. For spherical particles, pressure is a state function, independent of the wall potential \cite{Solon2015,Solon2015PRL}. Furthermore, the normal pressure applied to a curved wall is on average the same as the pressure on a flat wall \cite{Nikola2016}. Active particle ratchets, in which the force applied by active particles to objects is used to perform work, have been implemented both in experiments and simulations \cite{Reichhardt2017,Bechinger2016,Angelani2009,Angelani2011,DiLeonardo2010,Galajda2007,Kaiser2014,Wan2008}. Work can be extracted from such systems, because they are intrinsically out of equilibrium, due to self-propelled particles constantly inserting energy into the system. However, not every non-equilibrium system can be used to extract work. In fact, it is necessary to break both time reversal and space inversion symmetries in order to do so \cite{Tailleur2009,Magnasco1993,Julicher1997,Feynman1966}. Time reversal symmetry in these systems is broken by the interactions between the active particles and the objects. In previously studied active particle ratchets \cite{Reichhardt2017,Bechinger2016,Angelani2009,Angelani2011,DiLeonardo2010,Galajda2007,Kaiser2014,Wan2008}, space inversion is broken by the geometry of the moving objects or their environment, while the particle activity is uniform.

In this work we extend previous studies of the pressure in dry active systems \cite{Marchetti2013rev} to systems with an activity gradient.
We study the pressure exerted by non-interacting point-like active particles with an activity gradient. In this case, a symmetric inert object can be moved due to the pressure, since space inversion symmetry is broken by the gradient in the motion parameters.
This work is motivated by recent experiments \cite{Almonacid2015} which observed the motion of the oocyte (immature egg cell) nucleus from the cortex to the center, due to a gradient in the activity of vesicles.
Because of the long (hours) timescale of the nucleus motion and the possible lack of momentum conservation in the oocyte experiment \footnote{The vesicles are driven by molecular motors that walk on actin filaments. The actin network that fills the volume of the oocyte is dynamic, allowing the viscous motion of the nucleus on long time-scales. The momentum that is imparted by the active vesicles to the nucleus could be balanced by the forces that are transmitted through the actin network to the rigid cortex of the oocyte, and to the rest of the lab.}, hopefully the simple active particle model we studied can help understand the experiment.
We wish to find microscopic rules of motion for the vesicles, modeled as active particles, which explain the observed quantities: vesicle density and velocity profiles, and the resulting velocity of the nucleus.
Additionally, our work can describe synthetic systems in which activity can be spatially controlled \cite{Lozano2016,Palacci2013,Buttinoni2013}.

It was shown that for spherical active particles \cite{Solon2015, Solon2015PRL} pressure is a state function which depends on the bulk density and the motion parameters. Since the particle dynamics do not conserve momentum, a gradient in the motion parameters can cause a pressure gradient which exerts a force on an object immersed in the active fluid. We ask whether this force is simply given by an integral of a local pressure, which is a function of the particle motion parameters, over the object's surface. If so, for a constant pressure gradient the force is equal to the pressure gradient times the object's volume, analogously to the Archimedes principle (AP) for the buoyant force on an object submerged in a fluid under gravity.
We calculate the force on a passive body in the presence of a gradient of activity in one and two dimensions, and show how the AP needs to be modified in active fluids.

\emph{One dimensional model ---} Consider a 1D system of run-and-tumble particles, with position-dependent speed $v(x)$ and tumble rate $\alpha(x)$ \cite{Schnitzer1993}, which are confined by hard walls in the domain $-d\le x \le d$. We assume that the walls have no effect on the orientation of the particles, as they do not experience torque \cite{Solon2015}. We neglect thermal diffusion (for a discussion of the effect of adding diffusion, see Appendix \ref{sec:archimedes}) and interactions between the particles, both for simplicity and because they have been found to be negligible in the biological system of interest (the oocyte \cite{Almonacid2015}). A symmetric object in such a system is a piston with identical surfaces on its two sides. To find the force the particles apply to the piston, we first calculate the density of the particles within the bounded domain, in the absence of a piston. The density is discontinuous at the walls, where a macroscopic number of particles accumulates \cite{Lee2013, Elgeti2015, Ezhilan2015}. We therefore write rate equations for the bulk densities of left and right moving particles $L(x,t)$ and $R(x,t)$ \cite{Schnitzer1993,Tailleur2008,Tailleur2009}, in addition to coupled equations for the numbers of particles accumulated on the walls \cite{comment1}:
\begin{equation} \label{eq:FP_v_x_Dr_x}
\begin{array}{ll}
\vspace{1mm}
\partial_t R = -\partial_x(v(x)R) + \frac{\alpha(x)}{2}(L-R) \\
\vspace{1mm}
\partial_t L = \partial_x(v(x)L) + \frac{\alpha(x)}{2}(R-L) \\
\vspace{1mm}
\partial_t N_L^{-d} = -J_L(-d)-\frac{\alpha(-d)}{2} N_L^{-d} \\
\vspace{1mm}
\partial_t N_R^{-d} = -J_R(-d)+\frac{\alpha(-d)}{2} N_L^{-d} \\
\vspace{1mm}
\partial_t N_L^{d} = J_L(d)+\frac{\alpha(d)}{2} N_R^{d} \\
\partial_t N_R^{d} = J_R(d)-\frac{\alpha(d)}{2} N_R^{d} ,\\
\end{array}
\end{equation}
where $J_R = v(x) R$ and $J_L = -v(x) L$ are the currents of right and left moving particles, and $N_{L/R}^x$ is the number of left/right-moving particles at the boundary position $x=\pm d$. Note that the number of particles accumulated on a wall that are moving away from it is zero (i.e. $N_R^{-d}=N_L^d=0$).
We set the total particle number to $N$, i.e. $\int_{-d}^d\rho(x) \mathrm{d}x + N_L^{-d}+N_R^d = N$, where $\rho(x)=R(x)+L(x)$ is the total particle density. Under this constraint, the steady state solution of Eq.~\ref{eq:FP_v_x_Dr_x} is
\begin{equation} \label{eq:FP_sol}
\begin{array}{ll}
\vspace{1mm}
\rho(x)=\frac{c}{v(x)}, \; c=N (\int_{-d}^d\frac{1}{v(x)} \mathrm{d}x + \frac{1}{\alpha(-d)}+ \frac{1}{\alpha(d)})^{-1} \\
\vspace{1mm}
R(x)=L(x)=\frac{1}{2}\rho(x) \\
N_L^{-d}=\frac{c}{\alpha(-d)} , \; N_R^d=\frac{c}{\alpha(d)} , \; N_R^{-d}=N_L^d=0
\end{array}
\end{equation}
As shown in \cite{Schnitzer1993} for any dimension, the bulk density is inversely proportional to $v$, and is independent of the \emph{local} $\alpha$. The wall accumulation depends on the local value of $\alpha$, but not on the \emph{local} $v$.

The force on each of the walls can be calculated using the density. For example, the force on the right wall is given by the number of accumulated particles moving against the wall $N_R^d$, multiplied by the force a single particle applies, $F_1 = \mu_t^{-1} v$, where $\mu_t$ is the translational mobility: $F = N_R^d F_1 =c\mu_t^{-1} \ell_p(d)$, where $\ell_p\equiv v/\alpha$ is the persistence length.

\begin{figure}[t]
\includegraphics[width=1\linewidth]{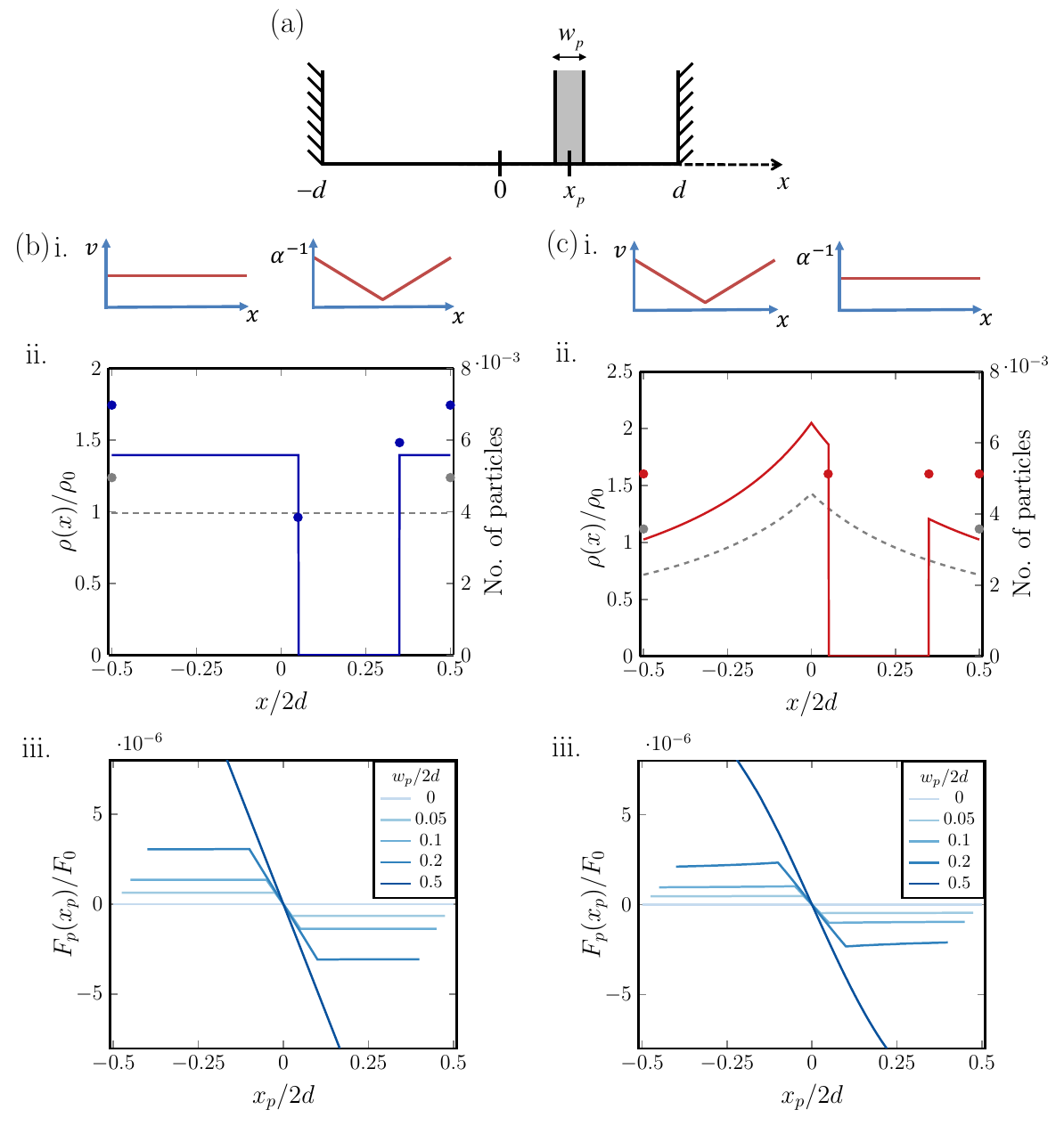}
\caption{(a) The 1D system (b) For $v=\frac{d}{200\tau_0}$, and $\alpha^{-1}(x)=\tau_0(|x|/d+1)$: i. $v$, $\alpha$ sketch ii. particle density (lines) and number of accumulated particles on the edges (dots), with (colored, solid) and without (gray, dashed) the piston ($x_p/2d=0.2$, $w_p/2d=0.3$). iii. The force on the piston for varying values of $w_p$.
(c) For $\alpha^{-1}=2\tau_0$ and $v(x)=\frac{|x|+d}{400\tau_0}$: i-iii as in (b). The force is directed towards the center, and its magnitude saturates when the entire piston is in one side of the system. Note that $\ell_p=v/\alpha$ is the same in (b) and (c). ($\rho_0\equiv \frac{N}{2d}$, $F_0\equiv\frac{2dN}{\mu_t\tau_0}$)}
\label{fig:1D}
\end{figure}

Next we find the force applied to an immersed piston with hard wall edges.
We denote the position of the piston's center by $x_p$, the piston's width by $w_p$, and the positions of the left and right edges of the piston by $x_p^{l/r}=x_p \mp w_p/2$  (Fig.~\ref{fig:1D}(a)). Since each of the two parts into which the piston divides the system is itself a 1D box like the one we solved above, the steady-state density of particles in each part is given by Eq.~\ref{eq:FP_sol}, where we replace the normalization constant and the total number of particles for the left/right sides by $c_{1/2}$ and $N_{1/2}$, respectively. The force on the piston is thus given by
\begin{equation} \label{eq: v_x_Dr_x_Fp}
F_p = N_R^{x_p^l}F_1(x_p^l) - N_L^{x_p^r}F_1(x_p^r)
    = \mu_t^{-1} \left(c_1 \ell_p(x_p^l) - c_2 \ell_p(x_p^r)  \right)
\end{equation}
The piston divides the system into two disconnected parts, and the force on it depends on the number of particles in each part, which determines the values of $c_{1,2}$. We choose to set $c_1=c_2$, as would happen in the large system size limit where the density profile is not affected by the insertion of an object. It is also true for periodic boundary conditions, and in higher dimensions for objects that do not divide the system into disconnected parts.
This choice, along with particle conservation, $N_1+N_2=N$, determine $N_1$ and $N_2$.

Setting $c_1=c_2$ in Eq.~\ref{eq: v_x_Dr_x_Fp} yields a force on the piston in the direction of the edge with the smaller local $\ell_p$. Thus a gradient in $\ell_p$ is necessary in order to move the piston. The force on the piston satisfies the AP: it is a sum over the piston edges of the local pressure $P=c_1\mu_t^{-1}\ell_p(x)$. Hence for a constant $\partial_x\ell_p$, the pressure gradient is constant and the force is equal to the pressure gradient times the piston width: $F_p=-\partial_xP w_p$.

The force on each edge of the piston is proportional to the local $\ell_p$ since it is the product of the number of particles accumulated on the piston edge, which is $\propto 1/\alpha$, and the force exerted by each particle, which is $\propto v$. While both of these components can be spatially dependent, it is easier to understand the limiting cases where either $v$ or $\alpha$ is constant and the other has spatial dependence.
In the case of a constant $v$ and a spatially dependent $\alpha$, the bulk density is constant, and the force on the piston pushes it towards small $\alpha$ regions (Fig.~\ref{fig:1D}(b)).
In the case of a constant $\alpha$ but varying $v$, the bulk density is nonuniform, and the particle accumulation on all surfaces is the same due to the constant $\alpha$. The force on the piston is proportional to the difference in $v$ between the two edges, pushing the piston towards the direction of smaller $v$ (Fig.~\ref{fig:1D}(c)).

\emph{Two dimensional model ---} Consider a two dimensional system of point-like non-interacting active particles with position dependent speed $v$, rotational diffusion rate $D_r$ and tumble rate $\alpha$.
We begin by studying the motion of a passive disk inside a circular domain (Fig.~\ref{fig:2D_sim}), using simulations (for simulation details, see Appendix \ref{sec:sim}) to examine the two limiting cases where either the speed $v$ or the persistence time $\tau=(\alpha+D_r)^{-1}$ varies spatially, while the other is constant. The particle density shows similar behavior to 1D: the bulk density is $\rho\propto v^{-1}$ \cite{Schnitzer1993,Cates2015} (Fig.~\ref{fig:2D_sim}(c)), while the accumulation on the edge of the disk depends on the local $\tau$ but not on the local $v$ (Fig.~\ref{fig:2D_sim}(d)).

\begin{figure}[h]
\includegraphics[width=1\linewidth]{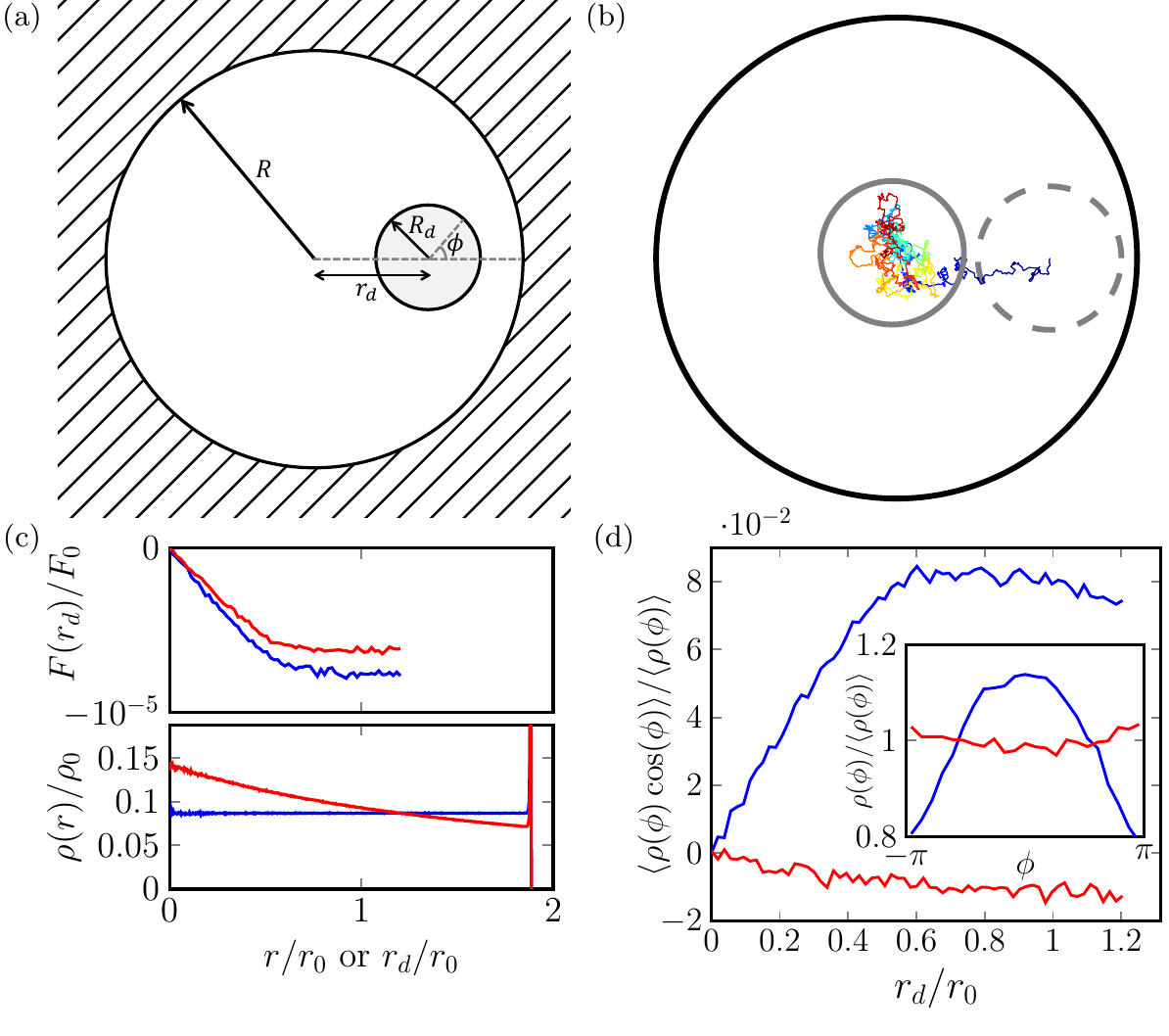}
  \caption[]{(a) The 2D system of a disk inside a circular box of active particles used in (b-d). (b) The trajectory of a disk in a $D_r$ gradient, where the line color represents time. The initial and final positions of the disk are marked by dashed and solid gray circles, respectively ($v=\frac{r_0}{30\sqrt{\pi}\tau_0}$, $D_r^{-1}=\tau_0(\frac{r}{R}+1)$, $\mu_t^{particle}/\mu_t^{disk}=30$). In (c-d), blue corresponds to $v$ and $D_r$ as in (b), and red to $v=\frac{r_0}{60\sqrt{\pi}\tau_0}(\frac{r}{R}+1)$, $D_r^{-1}=2\tau_0$. (c) The particle density as a function of the radial coordinate without a disk (bottom), and the average force on a disk as a function of $r_d$ (top).
The force is time averaged for a static disk. The validity of this for a moving disk is discussed in Appendix \ref{sec:static_vs_moving}.
 (d) The normalized angular average over $\rho(\phi)\cos(\phi)$ in a narrow ring around the disk, as a function of $r_d$. This quantifies the asymmetry in the particle accumulation on the disk surface. Inset: The density in the narrow ring as a function of the angle $\phi$ ($r_d=\frac{4 r_0}{3\sqrt{\pi}}$). For a gradient in $D_r$, more particles are accumulated on the side with smaller $D_r$, while for a varying $v$ the accumulation on the disk edge is nearly uniform. ($R=\frac{10 r_0}{3\sqrt{\pi}}$, $\alpha=0$, $r_0 \equiv \sqrt{\pi} R_d$, $\rho_0\equiv \frac{N}{r_0^2}$, $F_0\equiv\frac{r_0N}{\mu_t\tau_0}$)}
\label{fig:2D_sim}
\end{figure}

Unlike in 1D, finding the steady state density in 2D systems is a difficult problem \cite{Lee2013,Ezhilan2015,Elgeti2013,Elgeti2015,Fily2015}. However, using the method of \cite{Solon2015,Nikola2016}, it is possible to find the force on an object using the bulk density only, in the limit of small persistence length $\ell_p=\frac{v}{\alpha+D_r}$. When the persistence length grows, we show that two correction terms arise.

We start from the continuum equation for the distribution function $\mathcal{P}(\vec{r},\theta,t)$ of particles at position $\vec{r}$ and motility force direction $\theta$ at time t \cite{Solon2015,Nikola2016}:
\begin{equation} \label{eq:master_eq}
\partial_t \mathcal{P} = -\nabla \cdot [(v \hat{e}_{\theta} - \mu_t \nabla V)\mathcal{P}] + D_r \partial_{\theta}^2\mathcal{P} - \alpha \mathcal{P} + \frac{\alpha}{2 \pi}\int_0^{2 \pi} \mathrm{d}\theta'\mathcal{P}
\end{equation}
where $\hat{e}_{\theta}=(\cos \theta, \sin \theta)$ is the motility force direction, and $V(\vec{r})$ is the external potential due to surfaces such as the disk edge and system boundaries.

Consider an isolated object, with a narrow surface potential, inside the system. Following \cite{Solon2015,Nikola2016}, by using moments of Eq.~\ref{eq:master_eq}, and the fact that the force on the object is given by $\vec{F}^{tot} = \int_S \rho \nabla V  \mathrm{d}^2 \vec{r}$, where $S$ is an area containing the object and no other potentials, we obtain the following expression for the total force the particles exert on the object (see Appendix \ref{sec:2D_F}):
\begin{equation} \label{eq:F_tot}
\mu_t F^{tot}_x = -\vec{\mathcal{J}}_x - \int_{\partial S} \ell_p \vec{M}_{x1} \cdot \hat{n} \mathrm{d}\ell + \int_S(\nabla\ell_p) \cdot \vec{M}_{x1} \mathrm{d}^2\vec{r}
\end{equation}
where the axes were chosen such that $\vec{F}^{tot} \parallel \hat{x}$, $\hat{n}$ is a unit vector normal to $\partial S$, and $d\ell$ is a line element along $\partial S$. We define the integrated current $\vec{\mathcal{J}} \equiv \int_S \vec{J} \mathrm{d}^2\vec{r}$, where $\vec{J}$ is the current density, and $\vec{M}_{x1}=\frac{1}{2}v(\rho+m_{2x})\hat{x} + \frac{1}{2}vm_{2y} \hat{y} -\mu_t m_{1x} \nabla V$, with $\rho=\int_0^{2\pi} \mathrm{d}\theta \mathcal{P}(\vec{r},\theta)$, $m_{xn} = \int_0^{2\pi} \mathrm{d}\theta \mathcal{P}(\vec{r},\theta)\cos(n\theta)$ and $m_{yn} = \int_0^{2\pi} \mathrm{d}\theta \mathcal{P}(\vec{r},\theta)\sin(n\theta)$. 

Eq.~\ref{eq:F_tot} shows that the force can be divided into three terms, which we denote from left to right $F_J$, $F_{I1}$, and $F_{I2}$. The integrated current term $F_J$ is the only possibly non zero term in the case of constant parameters \cite{Nikola2016}. In this case, for a symmetric object such as a disk, $F=F_J=0$. The surface integral term $F_{I1}$ is the dominant term in the case of small $\ell_p$ with respect to the object lengthscales (see \ref{sec:scaling}). If $\partial S$ is in the bulk, $F_{I1}=\int_{\partial S} P \hat{x}\cdot\hat{n} \mathrm{d}\ell$ for the local pressure $P(\vec{r})=\frac{c}{2\mu_t}\ell_p(\vec{r})$, where $c=\rho(\vec{r})v(\vec{r})$ is a constant. Thus for a linear gradient in $\ell_p$, $F_{I1}$ has the form of the AP, i.e. it is proportional to $A_S$, the area of $S$, which is approximately the area of the object for $S$ along and slightly outside of the object surface, and independent of its shape: $\mu_t F_{I1} =  -\frac{c}{2}\frac{\partial\ell_p}{\partial x}A_S$. In 1D, this is the only non vanishing term (see Appendix \ref{sec:1D_F}).

\begin{figure}[ht!]
\includegraphics[width=1\linewidth]{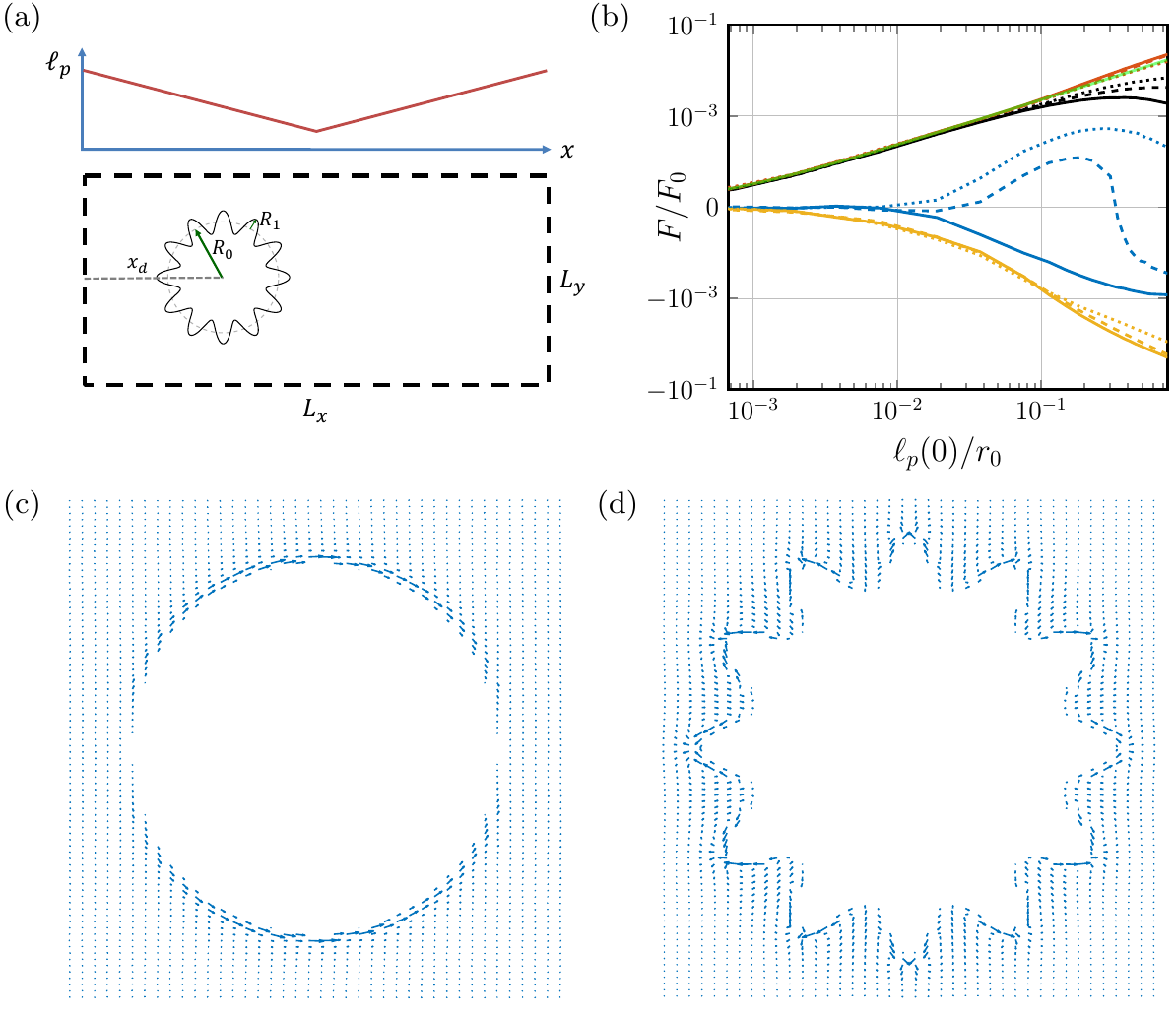}
  \caption[]{The effect of the persistence length and disk edge modulation. (a) Sketch of simulated system in b-d: A disk with a modulated edge inside a rectangular 2D system with periodic boundary conditions, and $\nabla \ell_p \parallel \hat{x}$. (b) The force on a modulated disk ($n=12$) as a function of $\ell_p(0)$, divided into 3 components, for a varying modulation amplitude: $R_1=$ $0$ (solid), $0.038 r_0$ (dashed), $0.075 r_0$ (dotted) and $S=$ an area concentric with the disk with edge at distance $r=R_d(\phi)+0.017r_0$ from the disk center. The modulated disk area $r_0^2$ is kept constant. Simulation results: the total force - black, $F_{I1}$ - red, $F_{I2}$ - yellow, $F_J$ - blue. The theoretical calculation of $F_{I1}$ (Eq.~\ref{eq:Archimedes_res}) is plotted in green. As the modulation amplitude increases, the total force on the disk is increased due to a reduction in magnitude of $F_{I2}$ and an increase in $F_J$ (F axis is transformed using a log-modulus of $10^5F/F_0$). (c,d) The current density around the disk for a modulation amplitude of 0 (c) and $0.075 r_0$ (d) for $\ell_p(0)=\frac{2r_0}{3\sqrt{\pi}}$. ($L_x=\frac{20r_0}{3\sqrt{\pi}}$, $L_y=\frac{10r_0}{3\sqrt{\pi}}$, $x_{d}=\frac{5r_0}{3\sqrt{\pi}}$, $F_0\equiv\frac{r_0N}{\mu_t\tau_0}$, $v=r_0/\tau_0$, $D_r^{-1}=\tau_0 \frac{\ell_p(0)}{r_0}(|\frac{x}{L_x}-0.5|+0.5)$, $\alpha=0$. Similar results are obtained for $\alpha^{-1}=\tau_0 \frac{\ell_p(0)}{r_0}(|\frac{x}{L_x}-0.5|+0.5)$, $D_r=0$, as shown in Appendix \ref{sec:Force_terms_Dr_vs_alpha}.)}
\label{fig:2D_pbc}
\end{figure}

In the small $\ell_p$ limit, the force is approximately equal to $F_{I1}$ and therefore it satisfies the AP. As $\ell_p$ grows, the current of particles slipping around the object grows and the integrated current $F_J$ and area integral term $F_{I2}$ grow and cause a deviation of the total force from $F_{I1}$ (Fig.~\ref{fig:2D_pbc}). For a disk, the two correction terms were observed in simulations to have opposite sign from $F_{I1}$, and thus reduce the total force (Fig.~\ref{fig:2D_pbc}(b)). Since the correction terms seem to be related to slippage of particles on the surface, we checked whether the force on the disk can be enhanced by reducing the slippage and decreasing these terms. We added a sinusoidal modulation to the disk radius: $R_d(\phi)=R_0+R_1 \cos(n\phi)$, where $\phi$ is the polar angle around the disk center. We keep $F_{I1}$ constant by using a linear $\ell_p$ gradient and keeping the area of the object constant, to isolate the effects of the other two terms. We find that indeed magnification of the total force can be achieved by modulation of the edge (Fig.~\ref{fig:2D_pbc}(b)). This is in accordance with previous studies, which showed that the curvature of the boundary affects the force applied by active particles in the case of uniform activity \cite{Fily2014,Harder2014,Mallory2014b,Smallenburg2015,Yan2015,Nikola2016}.

\emph{Discussion ---} We showed that while the AP applies in 1D, in 2D it generally needs to be modified and is valid only for a small persistence length. When the AP applies, the force exerted by the particles on an object is an integral over a local pressure that is proportional to the persistence length, which plays the role of the gravitational potential in the original AP. Thus this force tends to push objects in the direction of $-\nabla \ell_p$. This means that it is possible that the force on the container is nonzero, which is enabled by the breaking of momentum conservation by the particle motion dynamics.

Our model parameters and results can be linked to experimental measurements of particle trajectories, even if sampled at low frequency with respect to the persistence time $\tau$: knowing the long time mean squared displacement and the particle density is sufficient to determine $v(x)$ and $\tau(x)$ (see Appendix \ref{sec:exp_link}).
The force calculation for a passive object holds for a moving body in the limit where its velocity is slow with respect to the velocity of the active particles, such that the particles attain their steady state distribution throughout the motion. In addition, since the system is overdamped, the velocity of the object is proportional to the force exerted on it. Thus our force calculation can be used to obtain the velocity of slow moving objects.

Returning to the oocyte \cite{Almonacid2015}, the vesicle density was observed to be approximately uniform, while the measured vesicle velocity increases from the center to the edge of the cell. The resultant force on the nucleus pushes it towards the cell center. Since the sample time used for obtaining the vesicle positions was of the order of magnitude of $\tau$, the measured velocity depends both on the microscopic $v$ and $\tau$, making it possible that the observed spatial variation is due to a variation in either of them. The experimental results are consistent with our model for the case of a $\tau$ that increases from the center to the edge, and a constant $v$ (see Appendix \ref{sec:exp_data_analysis}), possibly indicating a spatial variation in the structure of the actin network.

\begin{acknowledgments}
\emph{Acknowledgements ---} We thank Marie-Helen Verlhac and Maria Almonacid for access to the experimental data. NR thanks Ada Yonath and the Kimmelman center for financial support. NSG and NR thank the support of the Schmidt Minerva Center. NSG is the incumbent of the Lee and William Abramowitz Professorial Chair of Biophysics and this research was supported by the ISF (Grant No. 580/12). This research is made possible in part by the generosity of the Harold Perlman family.
\end{acknowledgments}

\onecolumngrid
\appendix
\section{\label{sec:2D_F}Derivation of the force applied to an object in 2D}
The continuum equation for the distribution function $\mathcal{P}(\vec{r},\theta,t)$ of particles at position $\vec{r}$ and motility force direction $\theta$ at time t \cite{Solon2015,Nikola2016}

\begin{equation} \label{eq:master_eq_APPENDIX}
\partial_t \mathcal{P} = -\nabla \cdot [(v \hat{e}_{\theta} - \mu_t \nabla V)\mathcal{P}] + D_r \partial_{\theta}^2\mathcal{P} - \alpha \mathcal{P} + \frac{\alpha}{2 \pi}\int_0^{2 \pi}\mathrm{d}\theta'\mathcal{P}
\end{equation}

where $\hat{e}_{\theta}=(\cos \theta, \sin \theta)$ is the motility force direction, $\mu_t$ is the translational mobility and $V$ is the external potential.

Consider an object inside a system of active particles with position dependent $v$, $\alpha$ and $D_r$. Assume the object's surface potential is very narrow. Define as in \cite{Nikola2016}, $m_{xn} = \int_0^{2\pi}\mathrm{d}\theta \mathcal{P}(\vec{r},\theta)\cos(n\theta)$ and $m_{yn} = \int_0^{2\pi}\mathrm{d}\theta \mathcal{P}(\vec{r},\theta)\sin(n\theta)$. Integrating Eq.~\ref{eq:master_eq_APPENDIX} $\int_0^{2\pi}\mathrm{d}\theta$ and assuming steady state gives:

\begin{equation} \label{eq:J_def}
0 = \partial_t \rho = - \nabla \cdot \left( v m_{x1} \hat{x} + v m_{y1} \hat{y} - \mu_t \rho \nabla V \right) \equiv -\nabla \cdot J
\end{equation}

where $J$ is the current density. Multiplying Eq.~\ref{eq:master_eq_APPENDIX} by $\cos(\theta)$ / $\sin(\theta)$ and integrating $\int_0^{2\pi}d\theta$ gives:

\begin{align} \label{eq:2D_moments}
m_{x1} &= -\frac{1}{D_r+\alpha} \nabla \cdot \left( \frac{1}{2}v(\rho+m_{x2})\hat{x} + \frac{1}{2}vm_{y2} \hat{y} -\mu_t m_{x1} \nabla V \right) \equiv  -\frac{1}{D_r+\alpha} \nabla \cdot \vec{M}_{x1}\\
m_{y1} &= -\frac{1}{D_r+\alpha} \nabla \cdot \left( \frac{1}{2}v m_{y2}\hat{x} + \frac{1}{2}v(\rho-m_{x2}) \hat{y} -\mu_t m_{y1} \nabla V \right) \equiv  -\frac{1}{D_r+\alpha} \nabla \cdot \vec{M}_{y1} \nonumber
\end{align}

The total force on the object is given by

\begin{equation}
\vec{F}^{tot} = \int_S \rho \nabla V \mathrm{d}^2 \vec{r}
\end{equation}

where $V$ is the interaction potential of the object with the active particles, and the integral is over an area $S$ containing the object ($\nabla V=0$ outside of $S$).

We can relate the total force on an isolated object and the total current in an area that contains it as in \cite{Nikola2016}, by integrating the definition of the current density $J$ (Eq.~\ref{eq:J_def}):

\begin{equation} \label{eq:F_J_rel}
\vec{\mathcal{J}} \equiv \int_S \vec{J}\mathrm{d}^2\vec{r} = -\mu_t \vec{F}^{tot} + \int_S\left[vm_{x1}\hat{x} + vm_{y1}\hat{y} \right]\mathrm{d}^2\vec{r}
\end{equation}

While in the constant parameter case the last term on the right hand side of Eq.~\ref{eq:F_J_rel} vanishes, here it does not. Focus on this term:

\begin{align}
\vec{F}_I &\equiv \frac{1}{\mu_t} \int_S\left[vm_{x1}\hat{x} + vm_{y1}\hat{y} \right]\mathrm{d}^2\vec{r} \\
&= -\frac{1}{\mu_t} \int_S \ell_p \left( \nabla \cdot \vec{M}_{x1} \hat{x} + \nabla \cdot \vec{M}_{y1} \hat{y} \right) \mathrm{d}^2\vec{r} \nonumber
\end{align}

From now on, we choose the $\hat{x}$ axis to be along the direction of the total force for simplicity. Therefore we will consider the $\hat{x}$ component of the force terms. Using integration by parts, we obtain

\begin{equation} \label{eq:FI}
\mu_t F_{Ix} = -\int_S \ell_p \nabla \cdot \vec{M}_{x1} \mathrm{d}^2\vec{r} = -\int_{\partial S} \ell_p \vec{M}_{x1} \cdot \hat{n}\mathrm{d}\ell + \int_S(\nabla\ell_p) \cdot \vec{M}_{x1}\mathrm{d}^2\vec{r}
\end{equation}

where $\hat{n}$ is a unit vector normal to $\partial S$, and $d\ell$ is a line element along $\partial S$. Rearranging Eq.~\ref{eq:F_J_rel}, taking only its $\hat{x}$ component (since we assumed the $\hat{y}$ component of the force vanishes) and plugging in Eq.~\ref{eq:FI}, we obtain:
\begin{equation} \label{eq:F_tot_APPENDIX}
\mu_t F^{tot}_x = -\mathcal{J}_x - \int_{\partial S} \ell_p \vec{M}_{x1} \cdot \hat{n}\mathrm{d}\ell + \int_S(\nabla\ell_p) \cdot \vec{M}_{x1}\mathrm{d}^2\vec{r}
\end{equation}

where $F^{tot}_x$ is the x component of the total force on the object $\vec{F}^{tot}$, and $\mathcal{J}_x$ is the x component of the integrated current $\vec{\mathcal{J}}$.

\section{\label{sec:archimedes}Archimedes' principle derivation for 2D systems}
Consider the surface integral contribution to the force, which we denote $F_{I1}$:

\begin{equation} \label{eq:FI1}
\mu_t F_{I1} = -\int_{\partial S} \ell_p \vec{M}_{x1} \cdot \hat{n}\mathrm{d}\ell
\end{equation}

We choose the integration area $S$ to follow the edge of the object, surrounding the object at a small distance. Assuming that $\ell_p$ is small, bulk values of the probability moments in $\vec{M}_{x1}$ are obtained on $\partial S$: $\rho(\vec{r})=\frac{c}{v(\vec{r})}$ \cite{Schnitzer1993,Cates2015}, $m_{x2}=m_{y2}=0$. In addition, $\nabla V=0$ there, since the potential is assumed to be narrow and the integration region to be wider. Hence on the edge $\partial S$:

\begin{equation}
\vec{M}_{x1} = \frac{1}{2}v(\rho+m_{x2})\hat{x} + \frac{1}{2}vm_{y2} \hat{y} -\mu_t m_{x1} \nabla V = \frac{c}{2}\hat{x}
\end{equation}

Plugging this into Eq.~\ref{eq:FI1} and using the divergence theorem we obtain

\begin{equation}
\mu_t F_{I1} = -\frac{c}{2}\int_{\partial S}\ell_p \hat{x}\cdot \hat{n} \mathrm{d}\ell= -\frac{c}{2}\int_{S} \nabla \cdot(\ell_p \hat{x}) \mathrm{d}^2r = -\frac{c}{2}\int_{S} \frac{\partial\ell_p}{\partial x} \mathrm{d}^2r
\end{equation}

Note that $F_{I1}=\int_{\partial S} P \hat{x}\cdot\hat{n}\mathrm{d}\ell$ has the form of an integration of the local pressure $P(\vec{r})=\frac{c}{2\mu_t}\ell_p(\vec{r})$ over the edge of the object (projected on the $\hat{x}$ axis, since we chose to consider the $\hat{x}$ component of the force). Since it has this form, with $P$ that does not depend on the shape of $\partial S$, the divergence theorem immediately yields that as in the Archimedes principle for the buoyant force on an object immersed in a fluid, if the pressure gradient is constant then $F_{I1}$ is proportional only to the volume of the object and independent of its shape: for a linear gradient in $\ell_p$ (equivalently, a linear $P$ gradient), $\frac{\partial\ell_p}{\partial x} = \text{const}$ and thus

\begin{equation} \label{eq:Archimedes_res}
\mu_t F_{I1} = -\frac{c}{2}\frac{\partial\ell_p}{\partial x}\int_{S} \mathrm{d}^2r = -\frac{c}{2}\frac{\partial\ell_p}{\partial x}A
\end{equation}

where $A$ is the area of $S$, which is approximately equal to the area of the object. Hence $F_{I1}=-\partial_x P A$, analogously to the Archimedes principle. Interestingly, the pressure $P(\vec{r})=\frac{c}{2\mu_t}\ell_p(\vec{r})$ can be obtained from the pressure found in \cite{Solon2015PRL} to be exerted by particles with constant $v$ and $\tau$ on a flat wall by plugging in $v(\vec{r})$, $\tau(\vec{r})$ and the spatial dependence of the bulk density $\rho(\vec{r})=\frac{c}{v(\vec{r})}$.

In the small $\ell_p$ limit, the total force on the object $F^{tot}\approx F_{I1}$ and therefore it satisfies the AP. As $\ell_p$ grows, the $F_J$ and $F_{I2}$ terms grow and the AP is no longer satisfied by the total force, which no longer has the form of an integral over the object edge of a local pressure. In addition, for a large enough $\ell_p$, the bulk values of the moments $\rho$, $m_{xn}$ and $m_{yn}$ are no longer be obtained near the object surface due to the current created around the object, making the above calculation invalid.

Note that the division of the force on an object into three terms (Eq.~\ref{eq:F_tot}) is generally valid for any area $S$ that contains the object. However, the values of each of the three force terms depend on the choice of $S$. We showed here that a natural choice, which gives a special physical meaning to $F_{I1}$, is to have $S$ narrowly follow the edge of the object. When the persistence length is small, the boundary layer near the object edge, in which bulk values of the probability moments $\rho$, $m_{xn}$ and $m_{yn}$ are not attained, is narrow. Then a choice of $S$ that narrowly follows the object edge and yet remains in the bulk is possible, and the term $F_{I1}$ has a simple form, thanks to the known bulk values of the probability density moments, which has the physical meaning of being an equivalent to the Archimedes buoyant force, proportional to the object area.

A system with a small translational diffusion would behave qualitatively similar to one without diffusion (as discussed in \cite{Cates2015}). However, diffusion complicates the calculation of the force on an object in several ways.
If we include translational diffusion with constant diffusivity $D_t$ in our model, then the term $D_t\nabla^2\mathcal{P}(\vec{r},\theta,t)$ is added to the right hand side of Eq.~\ref{eq:master_eq_APPENDIX}. Following its contribution in the force calculation, we find that Eq.~\ref{eq:F_tot}  is satisfied, with an additional surface term on the right hand side: $D_t\int_{\partial S} \rho \hat{n}\cdot \hat{x} \mathrm{d}\ell$. In addition to this direct contribution of the diffusion to the force the particles exert, the diffusion affects our results in two ways. The first is that we no longer have an analytic solution for the bulk density, and hence we no longer have an analytic solution for $F_{I1}$ even when $\partial S$ is in the bulk. Second, the boundary layer next to the object edge in which the probability moments do not attain their bulk values becomes wider, having a width that grows with $D_t$. Thus while the division of the force on an object into terms as shown in Eq.~\ref{eq:F_tot} is still valid (with the addition of the term above), a natural choice of $\partial S$ which gives a clear physical meaning to $F_{I1}$ no longer exists.

\section{\label{sec:1D_F}Calculation of the force on a wall in 1D using the Solon method}
In the main text, the force on a wall in a 1D system was calculated by solving the equations for the particle density. We will now calculate this force using the method of \cite{Solon2015,Nikola2016}. We obtain the same result, and show that the $F_J$ and $F_{I2}$ contributions to the force vanish in this case.

We will calculate the pressure on the wall at $x=d$. We will use equations for the right and left moving particle densities, which are like those in main text Eq.~1, but with the wall potential written in the equation - instead of treating the wall as a boundary condition as was done in the main text. The equations are:
\begin{align} \label{eq:FP_Solon}
\partial_t R &= -\partial_x(v(x)R) + \frac{\alpha(x)}{2}(L-R) + \mu_t\partial_x(\partial_xV R) \\
\partial_t L &= \partial_x(v(x)L) + \frac{\alpha(x)}{2}(R-L) + \mu_t\partial_x(\partial_xV L) \nonumber
\end{align}

Summing and subtracting the equations gives:
\begin{align} \label{eq:rho_sigma}
\partial_t \rho &= -\partial_x \left( v \sigma - \mu_t \partial_x V \rho \right) = -\partial_x J \\
\partial_t \sigma &= -\partial_x(v \rho) - \alpha\sigma + \mu_t \partial_x(\partial_x V \sigma) \label{eq:rho_sigma2}
\end{align}

Now consider the steady state. Since Eq.~\ref{eq:rho_sigma} implies that in steady state $J=\text{constant}$, demanding $J=0$ on the wall gives that $J=0$ everywhere, and therefore $F_J=0$. From $J=0$ and Eq.~\ref{eq:rho_sigma} we obtain that $\rho \partial_x V = \frac{v}{mu_t}\sigma$. Using this, we get that the pressure on the wall at $x=d$ is:

\begin{equation}
P = \int_{d-\epsilon}^{d+\epsilon} \rho \partial_xV \mathrm{d}x = \int_{d-\epsilon}^{d+\epsilon}\frac{v}{\mu_t}\sigma \mathrm{d}x
\end{equation}

using Eq.~\ref{eq:rho_sigma2} we obtain that in steady state $\sigma = \frac{1}{\alpha} \partial_x \left(-v \rho + \mu_t \partial_x V \sigma)\right)$. Substituting this into the expression above for the pressure, we obtain:
\begin{align}
P &= \frac{1}{\mu_t} \int_{d-\epsilon}^{d+\epsilon}\mathrm{d}x \ell_p \partial_x\left(-v \rho + \mu_t \partial_x V \sigma \right) \\
 &= \frac{1}{\mu_t} \big(\ell_p (-v\rho + \mu_t \partial_x V \sigma) |_{d-\epsilon}^{d+\epsilon} - \int_{d-\epsilon}^{d+\epsilon}\mathrm{d}x \partial_x\ell_p(-v\rho + \mu_t\partial_xV \sigma) \big) \\
 &= \frac{c}{\mu_t}\ell_p(d-\epsilon) - \frac{1}{\mu_t} \int_{d-\epsilon}^{d+\epsilon} \mathrm{d}x \partial_x \ell_p (-v\rho + \mu_t \partial_x V \sigma)
\end{align}

Taking $\epsilon \to 0$, the first term ($F_{I1}$) gives the same result we got for the force on the wall from the exact solution. To calculate the second term ($F_{I2}$), we use the exact solution (main text Eq.~2):
\begin{align}
\rho(x) &= \frac{c}{v(x)} + \frac{c}{\alpha(d)} \delta(x-d) + \frac{c}{\alpha(-d)}\delta(x+d) \\
\sigma(x) &= \frac{c}{\alpha(d)} \delta(x-d) - \frac{c}{\alpha(-d)}\delta(x+d)
\end{align}

Plugging this into the 2nd term gives:
\begin{align}
&- \frac{1}{\mu_t} \int_{d-\epsilon}^{d+\epsilon} \mathrm{d}x \partial_x \ell_p(x) \Big( -c - c\frac{v(x)}{\alpha(d)}\delta(x-d) + \mu_t\partial_xV(x)\frac{c}{\alpha(d)}\delta(x-d) \Big) \\
&= -c\big(\ell_p(d+\epsilon)-\ell_p(d-\epsilon)\big) + c \partial_x\ell_p|_{x=d} \Big( -\frac{v(d)}{\alpha(d)} + \mu_t\partial_xV|_{x=d}\frac{1}{\alpha(d)} \Big)
\end{align}

The first term here vanishes as we take $\epsilon \to 0$.
Since the accumulation of particles happens at the point in the potential where the active force is equal to the force the potential applies, which we assume to be $x=d$ (since the potential is extremely narrow around that point): $v(d)=\mu_t\partial_xV(d)$. Thus the second term also vanishes and we get that this entire integral is equal to zero, giving: $P(d)=\frac{c}{\mu_t}\ell_p(d)$.

Note the same conclusion could be reached without using the full solution, by realizing that particles with active force pointing towards the wall accumulate on the wall, and thus there $\rho=\sigma=R\propto \delta(x-x_d)$. Combining this with a no flux demand on the wall $J=v\sigma-\mu_t\rho\partial_xV=0$ gives that the second term vanishes.

\begin{figure}[h!]
\includegraphics[width=0.8\linewidth]{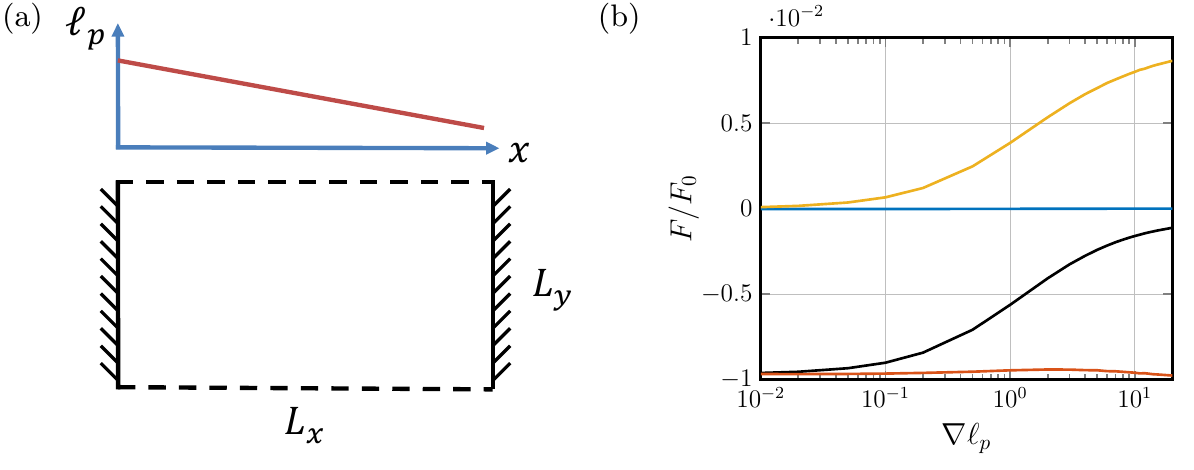}
  \caption[]{(a) system sketch of a 2D channel with walls in the $\hat{x}$ direction, and periodic boundary condition in the $y$ coordinate. The persistence length is a function of x only. (b) The force on the left wall divided into terms according to Eq.~5 (The total force - black, $F_{I1}$ - red, $F_{I2}$ - yellow, $F_J$ - blue) for the system in (a), for $S$ that is a narrow strip around the wall: $S=\{-\epsilon \le x \le 0.5x_0, 0 \le y \le L_y\}$, for some positive $\epsilon$. $F_J=0$ due to the translation symmetry in the $y$ direction. $F_{I2}$ is nonzero, contrary to the similar 1D system.
($L_x=50x_0$, $L_y=100x_0$, $v=x_0/\tau_0$, $D_r^{-1}=\tau_0(Bx/x_0+1)$, for varying $B=\nabla \ell_p$, $N=100$, $F_0\equiv \frac{x_0N}{\mu_t \tau_0}$).
}
\label{fig:2dchannel}
\end{figure}

To understand whether the correction term $F_{I2}$ is a basic result of the 2D motion the active particles, or it only appears for complex surface geometries, we consider a simple two dimensional system, with a wall parallel to the $\hat{y}$ axis and a $\ell_p$ gradient in the $\hat{x}$ direction (Fig.~\ref{fig:2dchannel}(a)). Assume that the system is invariant to translations in the $\hat{y}$ direction: the motion parameters and the wall interaction potential are functions of $x$ only.

Start from the continuum equation for the distribution function $\mathcal{P}(\vec{r},\theta,t)$ of particles at position $\vec{r}$ and motility force direction $\theta$ at time t (main text Eq.~4):

\begin{equation} \label{eq:master_eq11}
\partial_t \mathcal{P} = -\nabla \cdot [(v \hat{e}_{\theta} - \mu_t \nabla V(x))\mathcal{P}] + D_r \partial_{\theta}^2\mathcal{P} - \alpha \mathcal{P} + \frac{\alpha}{2 \pi}\int_0^{2 \pi}\mathrm{d}\theta'\mathcal{P}
\end{equation}

By taking moments of Eq.~\ref{eq:master_eq11}, and assuming steady state (all time derivatives vanish) we obtain (a private case of Eqs. 6,7 in \cite{Solon2015} SI):

\begin{align} \label{eq:moment_eqs}
0 &= -\partial_x(vm_{x1} -\mu_t\rho \partial_xV) \equiv \partial_x J\\
m_1 &= \frac{-1}{D_r+\alpha} \partial_x \Big( v\frac{\rho+m_{x2}}{2} - \mu_t m_{x1} \partial_x V \Big)
\end{align}


We would like to calculate the pressure exerted by the particles on a wall at $x=x_w$, assuming the interaction potential of the wall with the particles $V(x)$ diverges from $0$ to $\infty$ within $x_w-\epsilon<x<x_w$, and vanishes elsewhere. The regime $x>x_w$ is beyond the wall and contains no particles. The pressure on the wall is thus given by

\begin{equation} \label{eq:pressure_on_wall}
P = \int_{x_w-\epsilon}^{x_w+\epsilon} \rho(x)\partial_x V(x)\mathrm{d}x
\end{equation}

Using Eqs.~\ref{eq:moment_eqs} and the fact that $J=0$ ($J_y= 0$ due to the symmetry to translation in $y$, and $J_x=0$ on the wall since the particles cannot penetrate it), we obtain:

\begin{equation}
P = -\frac{1}{\mu_t} \int_{x_w-\epsilon}^{x_w+\epsilon} \frac{v}{D_r+\alpha} \partial_x \left( v\frac{\rho+m_{x2}}{2} - \mu_t m_{x1} \partial_x V \right)\mathrm{d}x
\end{equation}

Note that generally, this integral depends on the shape of $\partial_x V$. Performing integration by parts gives:

\begin{equation}
P = -\frac{1}{\mu_t} \ell_p\left( v\frac{\rho+m_{x2}}{2} - \mu_t m_{x1} \partial_x V \right)\Big|_{x=x_w-\epsilon}^{x=x_w+\epsilon} + \frac{1}{\mu_t}\int_{x_w-\epsilon}^{x_w+\epsilon}\partial_x\ell_p\left( v\frac{\rho+m_{x2}}{2} - \mu_t m_{x1} \partial_x V \right)\mathrm{d}x
\end{equation}

To relate this expression to our previous notations, note that the first term on the right hand side is $P_{I1} = F_{I1}/L_y$, and the second term is $P_{I2} = F_{I2}/L_y$. To calculate $P_{I1}$, we use the fact that at $x=x_w+\epsilon$ there are no particles and therefore $\rho=m_{xn}=0$.
Assuming the wall potential is very steep, already at $x_w-\epsilon$ for a small $\epsilon$ the moments of the probability density have their bulk values: $\rho = \frac{c}{v(x)}$ \cite{Schnitzer1993} ($c$ is a constant determined by the conservation of particle number equation $\int_{system}\rho(\vec{r})\mathrm{d}^2\vec{r}=N$), $m_{x1} = m_{x2} = 0$ (as we observed in the 1D analytic solution and in 2D simulations). Taking $\epsilon \to 0$, we obtain:

\begin{equation}
P_{I1} = \frac{c}{2 \mu_t} \ell_p(x_w)
\end{equation}

In the limit where the lengthscale over which the potential varies $\epsilon$ is much smaller than  the lengthscale over which the persistence length $\ell_p(x) \equiv \frac{v}{D_r + \alpha}$ varies $\ell_p \gg \epsilon \partial_x\ell_p$, the second term $P_{I2}$ is negligible with respect to $P_{I1}$. Thus the pressure is $P=P_{I1}$, and is independent of the wall potential $V(x)$.

However, the $P_{I2}$ term doesn't identically vanish as in the 1D wall case. The $J_x=0$ demand on the wall results in $\partial_x V=\frac{vm_{x1}}{\mu_t \rho}$. But now, differently from the 1D wall case, particles moving against the wall  have $\theta$ values in the entire half circle that has a component towards the wall, and thus $\rho\neq m_{x1}$ on the wall. In addition, contrary to 1D, there is a 1/2 in front of the $\rho v$ preventing the cancellation. Overall, we get that in the semi-1D geometry of a flat wall with an $\ell_p$ gradient perpendicular to the wall, $F_J=0$ due to symmetry but $F_{I2}$ is nonzero (Fig.~\ref{fig:2dchannel}).

\section{\label{sec:scaling} Scaling of the force terms}
Consider the force on a disk in a 2D bath of active particles (main text Eq.~5):
\begin{equation} \label{eq:F_tot2}
\mu_t \vec{F}^{tot}_x = \underbrace{-\vec{\mathcal{J}}_x}_{\equiv \mu_t F_J} + \underbrace{\int_{\partial S} \ell_p \vec{M}_{x1} \cdot \hat{n}\mathrm{d}\ell}_{\equiv \mu_t F_{I1}} - \underbrace{\int_S(\nabla\ell_p)\vec{M}_{x1}\mathrm{d}^2\vec{r}}_{\equiv \mu_t F_{I2}}
\end{equation}

We will use scaling arguments to show that if $S$ is a circle concentric with the disk, with radius $R_d+\epsilon$ for a small $\epsilon$, then in the limit of small persistence length with respect to the disk radius $\ell_p \ll R_d$, the integrated current term $F_J$ and $F_{I2}$ are negligible with respect to $F_{I1}$. First, the scaling of the $F_{I1}$ term:
\begin{equation} \label{eq:I1_scaling}
\mu_t F_{I1} = \int_{\partial S} \ell_p \vec{M}_{x1} \cdot \hat{n}\mathrm{d}\ell \propto c \int_{S} \nabla \ell_p \mathrm{d}^2\vec{r} \propto c \nabla \ell_p R_d^2
\end{equation}

where we assumed that $\partial S$ is in the bulk, and thus on it $\vec{M}_{x1} \propto \rho v = c$, and used the divergence theorem. We then used the fact that the area $S$ surrounding the disk is $\propto R_d^2$.

The scaling of the $F_{I2}$ term:
\begin{equation} \label{eq:I2_scaling}
\mu_t F_{I2} = \int_S(\nabla\ell_p)\vec{M}_{x1}\mathrm{d}^2\vec{r} \propto \nabla \ell_p \int_S \rho_{boundary} v\mathrm{d}^2\vec{r}
\end{equation}

where $\rho_{boundary}$ is the particle density in a narrow edge layer, of width $\epsilon$, near the disk's surface. To find how $\rho_{boundary}$ scales, since we are assuming a steady state, demand that the flux of particles into the boundary layer is equal to the flux out. The flux per unit length into the boundary layer is $j_{in}=\rho_{bulk} v = c$, while the flux per unit length out is $j_{out}=\frac{\rho_{boundary} \epsilon}{\tau}$. Demanding $j_{in}=j_{out}$ gives that $\rho_{boundary}=\frac{c \tau}{\epsilon}$. Substituting this into the scaling of $F_{I2}$, and using the fact that the part of $S$ in which the particle density is nonzero is a ring around the disk surface, with area proportional to $R_d \epsilon$, gives
\begin{equation} \label{eq:I2_scaling2}
\mu_t F_{I2} \propto \nabla \ell_p \frac{c \tau}{\epsilon} v R_d \epsilon \propto c \nabla \ell_p R_d \ell_p
\end{equation}

We got that $\frac{F_{I1}}{F_{I2}} \propto \frac{R_d}{\ell_p}$. Hence as long as $R_d \gg \ell_p$, the contribution of the $F_{I2}$ term to the force on the disk is negligible with respect to the contribution of the $F_{I1}$ term.

The scaling of the total current is different depending on how the persistence length $\ell_p$ compares to the disk radius $R_d$. In the regime of small persistence length $\ell_p \ll R$, transport around the disk is diffusive. Hence the scaling of the current density is
\begin{equation} \label{eq:J_scaling}
J \propto \nabla(D \rho_{boundary}) \propto \nabla(\frac{c}{\epsilon}v^2\tau^2) \propto \frac{c}{\epsilon}\nabla \ell_p^2 \propto \frac{c}{\epsilon} \ell_p \nabla \ell_p
\end{equation}

where in the first transition we used the diffusivity $D\propto v^2 \tau$ and the previously derived scaling of the boundary density $\rho_{boundary} \propto \frac{C \tau}{\epsilon}$.
Therefore the scaling of the total current in the ring is

\begin{equation} \label{eq:total_J_scaling}
\mathcal{J} = \int_S J \mathrm{d}^2 \vec{r} \propto \epsilon R_d J \propto c \ell_p \nabla \ell_p R_d
\end{equation}

meaning that in this case $F_J=-\mu_t\mathcal{J}$ scales like $F_{I2}$.

If the persistence length is of the order of magnitude of the disk radius or larger, $\ell_p \ge R_d$, the particle current around the disk is due to advective transport. Then the current density scaling is

\begin{equation} \label{eq:J_scaling2}
J \propto \Delta (v \rho_{boundary}) \propto \frac{c}{\epsilon}\Delta \ell_p \propto \frac{c}{\epsilon} R_d \nabla \ell_p
\end{equation}

where in the first transition we used $\rho_{boundary} \propto \frac{c \tau}{\epsilon}$, and in the second transition we used that the difference in persistence length between opposing edges of the disk $\Delta \ell_p$ is proportional to the persistence length gradient times the disk lengthscale $\nabla \ell_p R_d$. Hence the scaling of the total current in this case is

\begin{equation} \label{eq:total_J_scaling2}
\mathcal{J} = \int_S J \mathrm{d}^2 \vec{r} \propto \epsilon R_d J \propto c R_d^2 \nabla \ell_p
\end{equation}

In this case, $F_J$ scales as $F_{I1}$.

\begin{figure}[h!]
\includegraphics[width=0.8\linewidth]{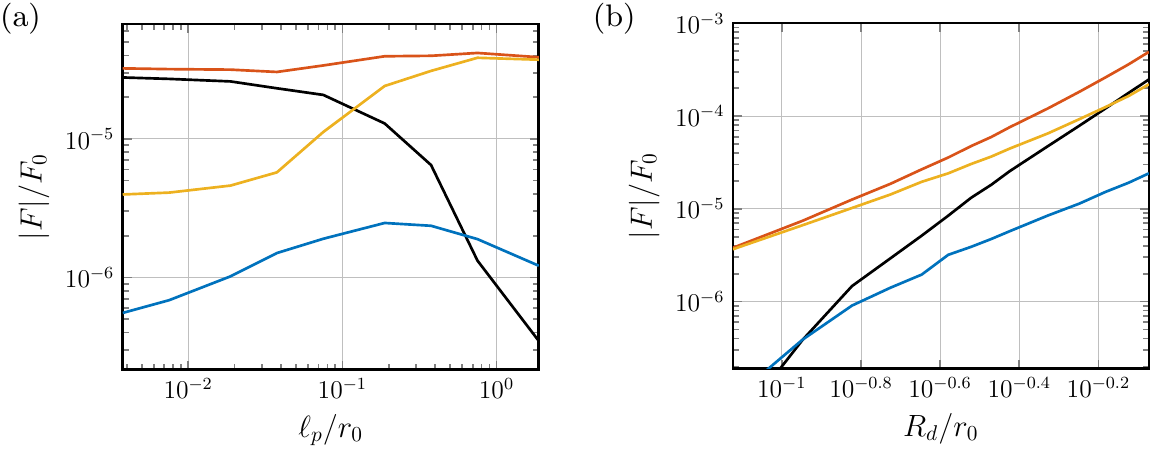}
  \caption[]{Scaling of the three terms of force on a disk, for: (a) $\ell_p$ changes at constant $R_d$ and $\nabla \ell_p$ (b) $R_d$ changes at constant $\ell_p$ and $\nabla \ell_p$ (The total force - black, $F_{I1}$ - red, $|F_{I2}|$ - yellow, $|F_J|$ - blue). Calculated in a rectangular system with periodic boundary conditions, as in Fig.~3, with $v=\frac{2r_0}{30\sqrt{\pi}\tau_0}$, in (a) $D_r^{-1}=\tau_0(|x/L_x-1/2|+0.1B)$ for various $B$ values, (b) $D_r^{-1}=\tau_0(5|x/L_x-1/2|+2.5)$.}
\label{fig:F_scaling}
\end{figure}

We test the scaling of the three force terms by observing how they vary in simulations as $\ell_p$ changes at constant $R_d$ and $\nabla \ell_p$ (Fig.~\ref{fig:F_scaling}(a)), and as $R_d$ changes at constant $\ell_p$ and $\nabla \ell_p$ (Fig.~\ref{fig:F_scaling}(b)). While the dependence of the terms on $R$ and $\ell_p$ is not with the exact power predicted by the scaling arguments, qualitatively the behavior is similar to the one predicted.
$F_{I1} \propto R_d^2$ is accurate, as expected by both the scaling argument and its exact calculation. $F_{I2}$ depends on $R_d$ with a weaker power, though it is larger than 1. It does grow as $\ell_p$ grows, but not linearly. The power with which $F_J$ depends on $R_d$ decreases, but not from 2 to 1 as predicted by the scaling arguments. As a function of $\ell_p$, $F_J$ has regimes of different behavior, but not of linear and constant growth.

\section{\label{sec:Force_terms_Dr_vs_alpha} Force terms for $D_r(x)$ vs $\alpha(x)$}
The two mechanisms included in the model (as defined by Eq.~\ref{eq:master_eq}) for a change of the direction of motion of a particle, which have corresponding change rates of $D_r$ and $\alpha$, explicitly appear in our expression for the force on an object in the same way, via the persistence length $\ell_p=v\tau$ with the persistence time $\tau=(D_r+\alpha)^{-1}$. However, they could have a different effect on the value of the force terms through the values of the probability moments $\rho$, $m_{xn}$ and $m_{yn}$, which appear in the expressions for $\vec{J}$ (Eq.~\ref{eq:J_def}) and $\vec{M_{x1}}$ (Eq.~\ref{eq:2D_moments}).

Run-and-tumble particles (RTP), which correspond to the limit case of $D_r=0$, and Active Brownian particles (ABP), which correspond to $\alpha=0$, have previously been shown to be quite similar. Their dynamics at the coarse grained diffusion-drift level is equivalent \cite{Cates2013ABPvsRTP,Solon2015ABPRTP}. Their accumulation near walls has been shown to be qualitatively similar, yet quantitatively different \cite{Elgeti2015}. Hence we expect the force terms in the two cases of RTP and ABP to be qualitatively similar, but differences in them could arise due to differences in the values of the probability moments $\rho$, $m_{xn}$ and $m_{yn}$, especially in the boundary region. In the bulk, in both cases (and for general $\alpha$ and $D_r$ values) $\rho(\vec{r})=\frac{c}{v(\vec{r})}$, $m_{xn}=0$ and $m_{yn}=0$, where the value of the constant $c$ can vary between the cases due to differences in the boundary behavior leading to a different number of particles being in the bulk. In the boundary region, greater differences in the moment values are expected.

\begin{figure}[h!]
\includegraphics[width=1\linewidth]{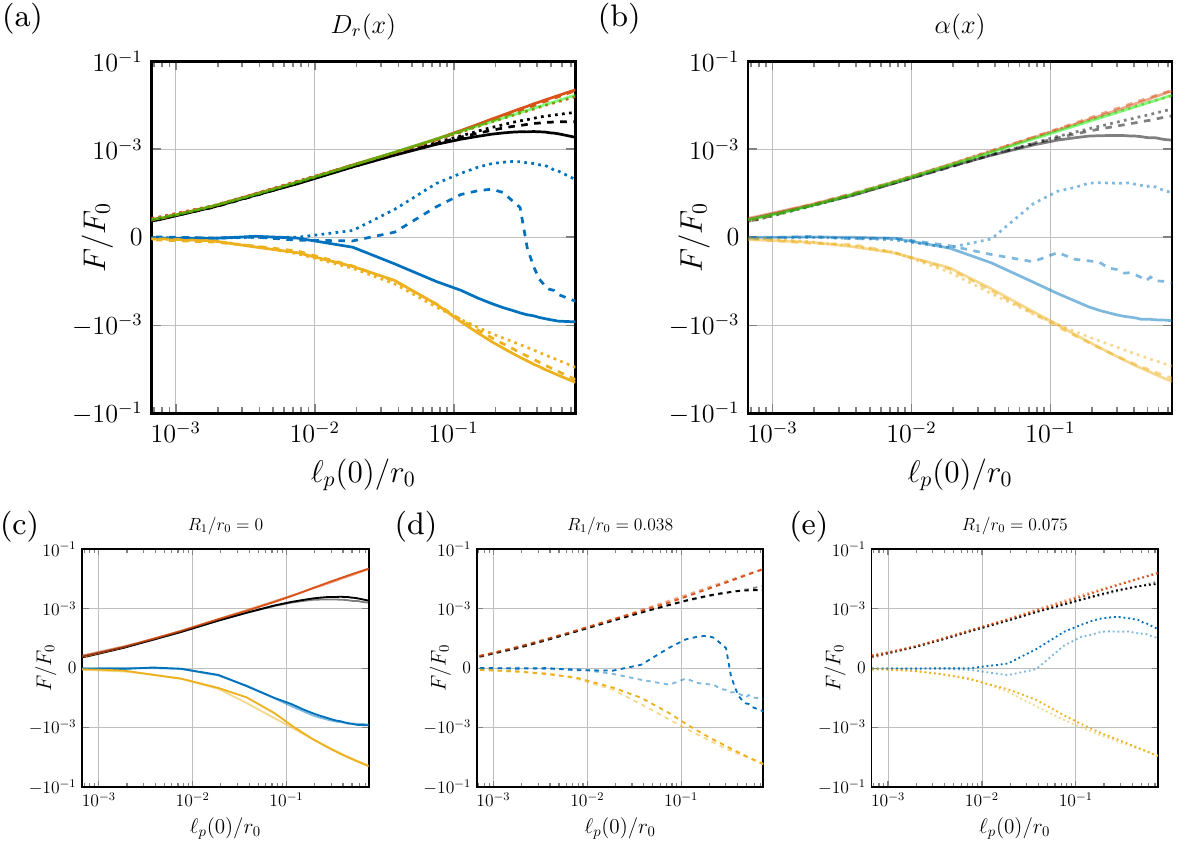}
  \caption[]{Force terms for $D_r(x)$ vs. $\alpha(x)$: (a) Same as Fig.~3(b), shows the force terms for a system with  $D_r^{-1}=\tau_0 \frac{\ell_p(0)}{r_0}(|\frac{x}{L_x}-0.5|+0.5)$ and
  $\alpha=0$ (b) The same plot as (a), for  $\alpha^{-1}=\tau_0 \frac{\ell_p(0)}{r_0}(|\frac{x}{L_x}-0.5|+0.5)$ and $D_r=0$. The persistence time $\tau=(D_r+\alpha)^{-1}$ is the same in (a) and (b). }
\label{fig:F_terms_Dr_vs_alpha}
\end{figure}

To test this, we consider a system with a persistence time $\tau(x)$ and compare the case of RTP: $\alpha^{-1}(x)=\tau(x)$ and $D_r=0$, to the case of ABP: $D_r^{-1}(x)=\tau(x)$ and $\alpha=0$, by plotting the force terms of Eq.~\ref{eq:F_tot} for the two cases from simulations (Fig.~\ref{fig:F_terms_Dr_vs_alpha}). We observe that for a small enough persistence length, the force terms have nearly identical values in the two cases, while for larger persistence length values there is a noticeable difference yet a qualitatively similar behavior. This makes sense because for small enough $\ell_p$ values, our chosen $\partial S$ is in the bulk, and the total force is approximately equal to the surface integral $F_{I1}$, which depends only on $\ell_p$ and the probability moments at $\partial S$ - in the bulk. The only possible difference is that $c$ could differ between the two cases due to different boundary behavior leading to a different number of particles in the bulk. However for the specific case we considered in Fig.~\ref{fig:F_terms_Dr_vs_alpha} there is no observable difference. As $\ell_p$ grows, the area integrals $F_J$ and $F_{I2}$ grow, and they include contributions from the boundary values of the probability moments, which are expected to be somewhat different between the two cases. Indeed we observe noticeable differences in their values between the cases.
In addition, at a large enough $\ell_p$, $\partial S$ is inside the boundary region and hence we expect to  also find differences between the two cases in the value of $F_{I1}$.

\section{\label{sec:sim} Simulation details}
We simulated 2D systems of non-interacting active Brownian particles by numerically integrating the overdamped Langevin equation of motion for each of the particles, using the Euler method. The equation of motion for each particle:
\begin{align}  \label{eq:2d_Langevin_EOM}
\partial_t \vec{r} &= v \hat{n}_{\theta}(t) + \mu_t \vec{F}^{ext} \\
\partial_t \theta &= \eta(t)
\end{align}

where  $v$ is the self propulsion speed, $\vec{r}=(x,y)$ is the particle's position, $\hat{n}_{\theta}=(\cos\theta,\sin\theta)$ is a unit vector in the direction of the motility force of the particle, and $\eta$ is white noise obeying the relations $\langle \eta(t) \rangle = 0$, $\langle \eta(t) \eta(t') \rangle = 2D_r \delta(t-t')$. $\mu_t$ is the mobility, and $\vec{F}^{ext}$ is the external force on the particle, due to interaction with the system boundaries and objects inside the system.

The system and disk boundaries were implemented by a force derived from narrow Lennard-Jones potentials truncated at the minimum, leaving just the repulsive part:

\begin{align}
V(\Delta r) &=
\begin{cases}
& 4 \epsilon \left( \left(\frac{\sigma}{\Delta r}\right)^{12} - \left(\frac{\sigma}{\Delta r}\right)^{6} \right) + \epsilon, \ \text{if}\ \Delta r < 2^{1/6}\sigma \\
& 0, \text{otherwise}
\end{cases}
\end{align}

where $\Delta r = r - r_{wall}$ is the particle position minus the wall position.

For a disk with a modulated edge, in polar coordinates around the disk center the force applied to a particle at position $(r,\phi)$ is derived from the potential above with $r_{wall} = R_d(\phi)=R_0+R_1 \cos(n\phi)$.

Additional details for simulations of the circular system presented in the main text Fig.~2:
The parameters of the interaction potential of the disk with the active particles were $\epsilon=7 \times 10^{-8} F_0 r_0$, $\sigma=\frac{1}{60\sqrt{\pi}}$. The number of particles was $N=100$.
(b): The disk trajectory shown is from a simulation with step size $dt=2 \times 10^{-2} \tau_0$, total simulation time of $4 \times 10^5 \tau_0$.
(c): $dt=2 \times 10^{-2} \tau_0$, total simulation time $1.6\times 10^8 \tau_0$. The particle density was obtained from samples every $100$ simulation time units, the force on the disk at each position is the average over all simulation steps.
(d): $dt=2 \times 10^{-2} \tau_0$, the density was measured in a ring around the disk of width $dr=0.01 r_0$, which is approximately the width of the repulsive truncated Lennard-Jones potential used for the interaction of the disk edge with the particles. The density was averaged from a total simulation time of $4\times 10^7 \tau_0$ for each disk position .

Additional details for simulations of the rectangular system with periodic boundary conditions presented in the main text Fig.~3:
The number of particles was $N=100$. The simulation step size was $dt=1.88 \times 10^{-4} \tau_0$. Each data point in (b) and the current density in (c) and (d) were obtained by binned averages over a simulation time of $1.5 \times 10^6 \tau_0$. The interaction potential parameters for the modulated disk were $\epsilon=3.7 \times 10^{-8} F_0 r_0$ and $\sigma=1.86\times 10^{-3} r_0$.
The average total force on the disk was directly calculated. $S$ with which we chose to work for the calculation of the force terms is a contour along the edge of the modulated disk at a distance $dr=\frac{r_0}{30\sqrt{\pi}}$ outside  it. We calculated binned averages of the particle density, current density, and the moments $m_{xn}$, $m_{yn} $ in two rings of width $dr$ around the modulated disk. $F_J$ was calculated by summing the current density inside $S$. $F_{I1}$ was calculated by using the density and moment values in the second ring as an estimate for the values on the edge of the first ring. $F_{I2}$ was calculated by subtracting $F_J$ and $F_{I1}$ from the total force. The theoretical calculation of $F_{I1}$ is Eq.~\ref{eq:Archimedes_res}, with the value of $c$ estimated by neglecting the wall accumulation of the particles.






\section{\label{sec:static_vs_moving} 2D static vs. moving disk simulations}
For our calculation of the force on a static object to be valid for a moving object, we must assume that the active particles move much faster than the object, and thus they are at steady state throughout the motion of the object at all times.

\begin{figure}[h!]
\includegraphics[width=0.4\linewidth]{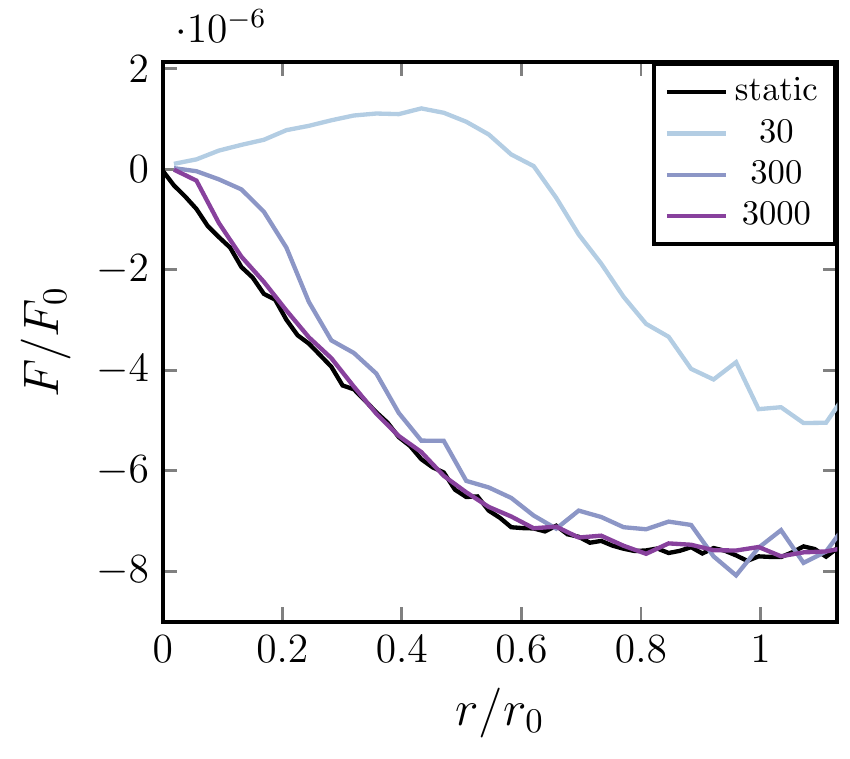}
  \caption[]{Force on a disk as a function of its position, for: 1. a static disk held in place (black), 2. a disk that can move due to the forces applied to it by the ABPs, for various ratios of particle and disk motilities $\mu_t^{particle}/\mu_t^{disk}=30,300,3000$ (colored lines). ($F_0\equiv\frac{r_0 N}{\mu_t\tau_0}$)}
\label{static_vs_dynamic}
\end{figure}

We check the validity of this assumption for a simulation with $v=\frac{r_0}{30\sqrt{\pi}\tau_0}$, $D_r^{-1}=\tau_0(\frac{r}{R}+1)$, $R=\frac{10r_0}{3\sqrt{\pi}}$, $R_d=r_0/\sqrt{\pi}$, $N=100$ in Fig.~\ref{static_vs_dynamic}. As the ratio of particle and disk mobilities $\mu_t^{particle}/\mu_t^{disk}$ grows, the static simulation better approximates the dynamic one.

\section{{\label{sec:exp_link}}Linking our model to experimental measurements} 
We can link the model parameters - the speed $v(\vec{r})$ and persistence time $\tau(\vec{r})=(\alpha(\vec{r})+(d-1)D_r(\vec{r}))^{-1}$ (in $d$ dimensions \cite{Cates2015}) - to  measurable quantities, which are extracted from low frequency imaging of the particle trajectories, with sample time $\Delta t \gg \tau$. First, the bulk density is related to the velocity profile by $\rho \propto v^{-1}$ \cite{Schnitzer1993}. The macroscopic velocity is measured from the sampled position differences $\Delta x$ over the time difference $\Delta t$: $v_m\equiv \Delta x / \Delta t$. Since $\Delta t \gg \tau$, $\Delta x\propto \sqrt{D\Delta t}$, where $D=v^2\tau/d$ is the diffusivity \cite{Cates2015}. Hence the spatial dependence of the macroscopic velocity is given by $v_m(\vec{r})\propto \sqrt{v(\vec{r})^2 \tau(\vec{r})}$, assuming the motion parameters vary spatially slowly enough ($v\nabla \tau \ll 1$, $\tau \nabla v \ll 1$) to make the diffusivity $D(\vec{r})=v(\vec{r})^2\tau(\vec{r})/d$ locally meaningful (this assumption is self-consistent for extracting the motion parameters of the oocyte vesicles, as seen from Fig.~\ref{fig:ABPparam3bins}). Note that while $v$ determines the shape of $\rho$ and vice versa, any $\rho$ and $v_m$ profiles can be obtained by tuning $v(r)$ and $\tau(r)$. This choice then determines the force on an object, which for small $\ell_p$ is in the direction of decreasing $\ell_p=v \tau$. For overdamped dynamics, this force is proportional to the object velocity, which is measurable by imaging \cite{Almonacid2015}.

Previously, in \cite{Almonacid2015}, the experimental measurements were used to propose that the active pressure is given by the following expression: $P \propto \rho(\vec{r})v_m({\vec{r}})^2$. From the results above we now see that this is simply $\propto \ell_p(\vec{r})$, and therefore gives the correct direction and spatial dependence of the active pressure, pushing in the direction of minimal $\ell_p$.


\section{\label{sec:exp_data_analysis}Experimental data analysis}
\begin{figure}[h!]
{\includegraphics[width=0.8 \linewidth]{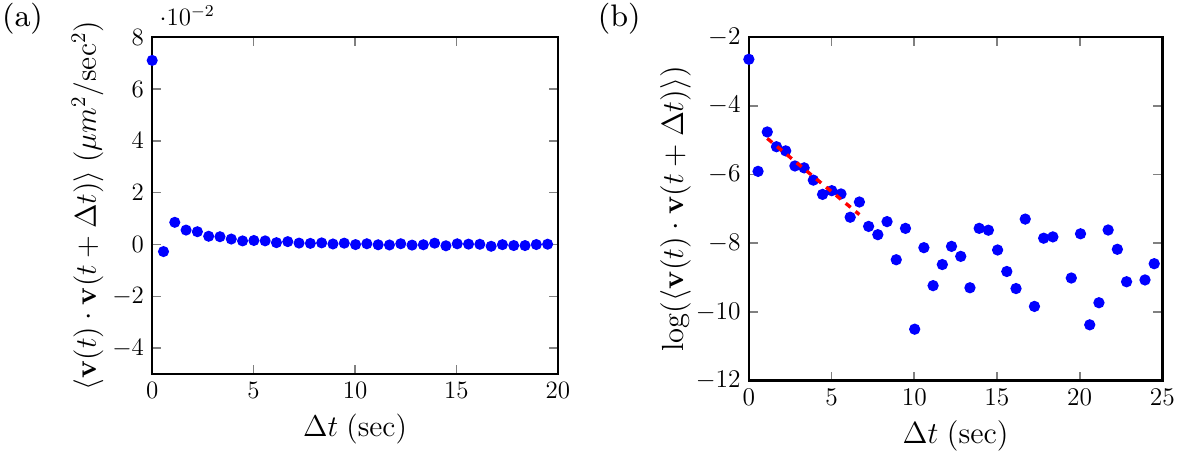}}
    \caption{Velocity-velocity correlation, in (a) linear scale, (b) log scale. The 2nd point at $\Delta t \approx 1/2$ sec is negative, but later a slight exponential decay can be observed. In (b) a linear fit (dashed red line) was performed for all points after the 2nd which have a value of more than $1\%$ of the initial correlation, out of which the parameters of an ABP model were extracted: $D_r\approx0.4 \; \mbox{sec}^{-1}$, $v\approx0.09 \; \mu m/\mbox{sec}$, $D\approx0009 \; \mu m^2/\mbox{sec}$.}
    \label{fig:vel-vel_corr}
\end{figure}

We calculated the velocity-velocity correlation function for the vesicle trajectories used in \cite{Almonacid2015}. We calculated $\langle \vec{v}(t+\Delta t)\cdot \vec{v}(t) \rangle$ where the averaging is over all vesicles' trajectories and over t, for $\Delta t$s that are integer multiplications of the sampling rate of the trajectories. The results are plotted in Fig.~\ref{fig:vel-vel_corr}. After a short timescale discrepancy at the first timestep after time zero, an exponential decay with time $\tau \approx 2.5\, \text{sec}$ is observed, in agreement with an active Brownian particle model.

The measured velocity of the vesicles, calculated from the experimental trajectory data, is a position difference divided by the sample time. Thus a discrete set of velocities was measured: $\vec{v}(k \Delta t)=\frac{\vec{\Delta x}(k \Delta t)}{\Delta t}$, where we denote $\vec{\Delta x}(k \Delta t)=\vec{x}(k \Delta t) - \vec{x}((k-1) \Delta t)$. In order to fit the measured velocity-velocity correlation by an ABP model, we calculated for an active Brownian particle with diffusion coefficient $D$, active velocity $v$ and rotational diffusion rate $D_r$:

\begin{equation}
\begin{array}{ll}
C(t\equiv k \Delta t) \equiv \langle \vec{v}(t'+k\Delta t)\cdot \vec{v}(t') \rangle_{t'} \\
\equiv \frac{1}{(k\Delta t)^2}\langle \vec{\Delta x}(t'+k \Delta t)\cdot \vec{\Delta x}(t') \rangle_{t'} \\
=\begin{cases}
    4(D+\frac{v^2}{2D_r})\Delta t + \frac{2v^2}{D_r^2}(e^{-D_r \Delta t}-1),& \text{if } k=1\\
    \frac{v^2}{D_r^2}(e^{D_r \Delta t}-1)^2e^{-D_rk\Delta t},              & \text{if } k>1
\end{cases}

\end{array}
\end{equation}

\begin{figure}[h!]
\includegraphics[width=0.4\linewidth]{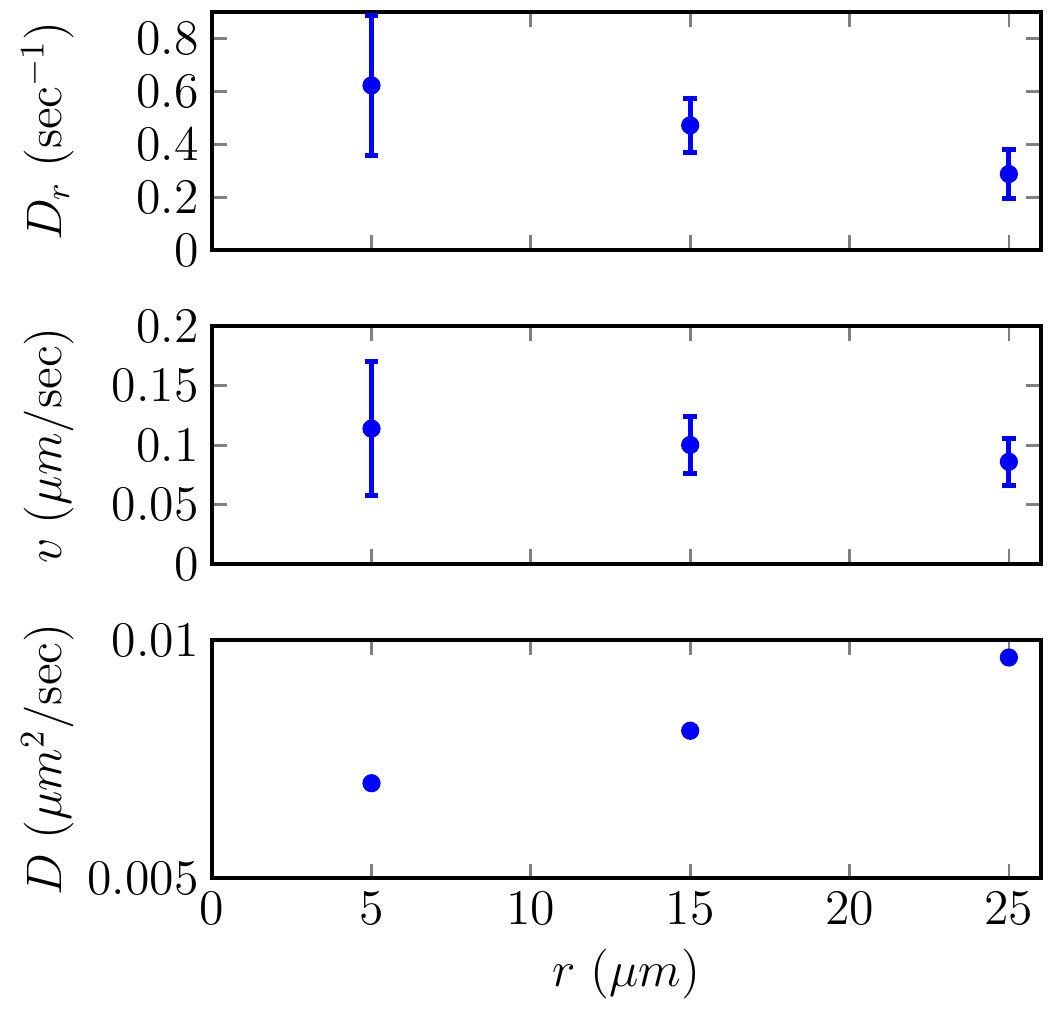}
  \caption[]{ABP model parameter values from fit to binned velocity-velocity correlation. Error bars are $95\%$ confidence bounds of the fit.}
\label{fig:ABPparam3bins}
\end{figure}

Separating the data into bins according to the mean distance from the cell center (r) of each trajectory, we obtained fits for the parameters of the ABP model. The results for a 3-bin division are shown in Fig.~\ref{fig:ABPparam3bins}. Though the data is noisy, it seems that $D_r$ decreases and $D$ increases towards the center, while $v$ is constant within the precision of our measurement.

From the fit of an ABP model to the velocity-velocity correlation, we find that $v\approx0.09 \, \mu m/sec$, and $\tau \approx 2.5 \, sec$ (Fig.~\ref{fig:vel-vel_corr}). Therefore the persistence length is $\ell_p=v\tau \approx 0.2 \, \mu m$, much smaller than the nucleus radius. Thus according to our active particle model, we expect the force on the nucleus to push towards the direction of minimal $\ell_p$, which is towards the center of the cell. This is in agreement with the observed motion of the nucleus from the cortex to the center.


\begin{thebibliography}{43}%
\makeatletter
\providecommand \@ifxundefined [1]{%
 \@ifx{#1\undefined}
}%
\providecommand \@ifnum [1]{%
 \ifnum #1\expandafter \@firstoftwo
 \else \expandafter \@secondoftwo
 \fi
}%
\providecommand \@ifx [1]{%
 \ifx #1\expandafter \@firstoftwo
 \else \expandafter \@secondoftwo
 \fi
}%
\providecommand \natexlab [1]{#1}%
\providecommand \enquote  [1]{``#1''}%
\providecommand \bibnamefont  [1]{#1}%
\providecommand \bibfnamefont [1]{#1}%
\providecommand \citenamefont [1]{#1}%
\providecommand \href@noop [0]{\@secondoftwo}%
\providecommand \href [0]{\begingroup \@sanitize@url \@href}%
\providecommand \@href[1]{\@@startlink{#1}\@@href}%
\providecommand \@@href[1]{\endgroup#1\@@endlink}%
\providecommand \@sanitize@url [0]{\catcode `\\12\catcode `\$12\catcode
  `\&12\catcode `\#12\catcode `\^12\catcode `\_12\catcode `\%12\relax}%
\providecommand \@@startlink[1]{}%
\providecommand \@@endlink[0]{}%
\providecommand \url  [0]{\begingroup\@sanitize@url \@url }%
\providecommand \@url [1]{\endgroup\@href {#1}{\urlprefix }}%
\providecommand \urlprefix  [0]{URL }%
\providecommand \Eprint [0]{\href }%
\providecommand \doibase [0]{http://dx.doi.org/}%
\providecommand \selectlanguage [0]{\@gobble}%
\providecommand \bibinfo  [0]{\@secondoftwo}%
\providecommand \bibfield  [0]{\@secondoftwo}%
\providecommand \translation [1]{[#1]}%
\providecommand \BibitemOpen [0]{}%
\providecommand \bibitemStop [0]{}%
\providecommand \bibitemNoStop [0]{.\EOS\space}%
\providecommand \EOS [0]{\spacefactor3000\relax}%
\providecommand \BibitemShut  [1]{\csname bibitem#1\endcsname}%
\let\auto@bib@innerbib\@empty
\bibitem [{\citenamefont {Solon}\ \emph
  {et~al.}(2015{\natexlab{a}})\citenamefont {Solon}, \citenamefont {Fily},
  \citenamefont {Baskaran}, \citenamefont {Cates}, \citenamefont {Kafri},
  \citenamefont {Kardar},\ and\ \citenamefont {Tailleur}}]{Solon2015}%
  \BibitemOpen
  \bibfield  {author} {\bibinfo {author} {\bibfnamefont {A.~P.}\ \bibnamefont
  {Solon}}, \bibinfo {author} {\bibfnamefont {Y.}~\bibnamefont {Fily}},
  \bibinfo {author} {\bibfnamefont {A.}~\bibnamefont {Baskaran}}, \bibinfo
  {author} {\bibfnamefont {M.~E.}\ \bibnamefont {Cates}}, \bibinfo {author}
  {\bibfnamefont {Y.}~\bibnamefont {Kafri}}, \bibinfo {author} {\bibfnamefont
  {M.}~\bibnamefont {Kardar}}, \ and\ \bibinfo {author} {\bibfnamefont
  {J.}~\bibnamefont {Tailleur}},\ }\href {\doibase 10.1038/nphys3377}
  {\bibfield  {journal} {\bibinfo  {journal} {Nature Physics}\ }\textbf
  {\bibinfo {volume} {11}},\ \bibinfo {pages} {673} (\bibinfo {year}
  {2015}{\natexlab{a}})},\ \Eprint {http://arxiv.org/abs/1412.3952v2}
  {arXiv:1412.3952v2} \BibitemShut {NoStop}%
\bibitem [{\citenamefont {Solon}\ \emph
  {et~al.}(2015{\natexlab{b}})\citenamefont {Solon}, \citenamefont
  {Stenhammar}, \citenamefont {Wittkowski}, \citenamefont {Kardar},
  \citenamefont {Kafri}, \citenamefont {Cates},\ and\ \citenamefont
  {Tailleur}}]{Solon2015PRL}%
  \BibitemOpen
  \bibfield  {author} {\bibinfo {author} {\bibfnamefont {A.~P.}\ \bibnamefont
  {Solon}}, \bibinfo {author} {\bibfnamefont {J.}~\bibnamefont {Stenhammar}},
  \bibinfo {author} {\bibfnamefont {R.}~\bibnamefont {Wittkowski}}, \bibinfo
  {author} {\bibfnamefont {M.}~\bibnamefont {Kardar}}, \bibinfo {author}
  {\bibfnamefont {Y.}~\bibnamefont {Kafri}}, \bibinfo {author} {\bibfnamefont
  {M.~E.}\ \bibnamefont {Cates}}, \ and\ \bibinfo {author} {\bibfnamefont
  {J.}~\bibnamefont {Tailleur}},\ }\href {\doibase
  10.1103/PhysRevLett.114.198301} {\bibfield  {journal} {\bibinfo  {journal}
  {Physical Review Letters}\ }\textbf {\bibinfo {volume} {114}},\ \bibinfo
  {pages} {1} (\bibinfo {year} {2015}{\natexlab{b}})},\ \Eprint
  {http://arxiv.org/abs/1412.5475} {arXiv:1412.5475} \BibitemShut {NoStop}%
\bibitem [{\citenamefont {Nikola}\ \emph {et~al.}(2016)\citenamefont {Nikola},
  \citenamefont {Solon}, \citenamefont {Kafri}, \citenamefont {Kardar},
  \citenamefont {Tailleur},\ and\ \citenamefont {Voituriez}}]{Nikola2016}%
  \BibitemOpen
  \bibfield  {author} {\bibinfo {author} {\bibfnamefont {N.}~\bibnamefont
  {Nikola}}, \bibinfo {author} {\bibfnamefont {A.~P.}\ \bibnamefont {Solon}},
  \bibinfo {author} {\bibfnamefont {Y.}~\bibnamefont {Kafri}}, \bibinfo
  {author} {\bibfnamefont {M.}~\bibnamefont {Kardar}}, \bibinfo {author}
  {\bibfnamefont {J.}~\bibnamefont {Tailleur}}, \ and\ \bibinfo {author}
  {\bibfnamefont {R.}~\bibnamefont {Voituriez}},\ }\href {\doibase
  10.1103/PhysRevLett.117.098001} {\ \textbf {\bibinfo {volume} {098001}},\
  \bibinfo {pages} {1} (\bibinfo {year} {2016})}\BibitemShut {NoStop}%
\bibitem [{\citenamefont {Marchetti}\ \emph {et~al.}(2016)\citenamefont
  {Marchetti}, \citenamefont {Fily}, \citenamefont {Henkes}, \citenamefont
  {Patch},\ and\ \citenamefont {Yllanes}}]{Marchetti201634}%
  \BibitemOpen
  \bibfield  {author} {\bibinfo {author} {\bibfnamefont {M.~C.}\ \bibnamefont
  {Marchetti}}, \bibinfo {author} {\bibfnamefont {Y.}~\bibnamefont {Fily}},
  \bibinfo {author} {\bibfnamefont {S.}~\bibnamefont {Henkes}}, \bibinfo
  {author} {\bibfnamefont {A.}~\bibnamefont {Patch}}, \ and\ \bibinfo {author}
  {\bibfnamefont {D.}~\bibnamefont {Yllanes}},\ }\href {\doibase
  http://dx.doi.org/10.1016/j.cocis.2016.01.003} {\bibfield  {journal}
  {\bibinfo  {journal} {Current Opinion in Colloid \& Interface Science}\
  }\textbf {\bibinfo {volume} {21}},\ \bibinfo {pages} {34} (\bibinfo {year}
  {2016})}\BibitemShut {NoStop}%
\bibitem [{\citenamefont {Takatori}\ \emph {et~al.}(2014)\citenamefont
  {Takatori}, \citenamefont {Yan},\ and\ \citenamefont {Brady}}]{Takatori2014}%
  \BibitemOpen
  \bibfield  {author} {\bibinfo {author} {\bibfnamefont {S.~C.}\ \bibnamefont
  {Takatori}}, \bibinfo {author} {\bibfnamefont {W.}~\bibnamefont {Yan}}, \
  and\ \bibinfo {author} {\bibfnamefont {J.~F.}\ \bibnamefont {Brady}},\ }\href
  {\doibase 10.1103/PhysRevLett.113.028103} {\bibfield  {journal} {\bibinfo
  {journal} {Phys. Rev. Lett.}\ }\textbf {\bibinfo {volume} {113}},\ \bibinfo
  {pages} {028103} (\bibinfo {year} {2014})}\BibitemShut {NoStop}%
\bibitem [{\citenamefont {Yang}\ \emph {et~al.}(2014)\citenamefont {Yang},
  \citenamefont {Manning},\ and\ \citenamefont {Marchetti}}]{Yang2014}%
  \BibitemOpen
  \bibfield  {author} {\bibinfo {author} {\bibfnamefont {X.}~\bibnamefont
  {Yang}}, \bibinfo {author} {\bibfnamefont {M.~L.}\ \bibnamefont {Manning}}, \
  and\ \bibinfo {author} {\bibfnamefont {M.~C.}\ \bibnamefont {Marchetti}},\
  }\href {\doibase 10.1039/C4SM00927D} {\bibfield  {journal} {\bibinfo
  {journal} {Soft Matter}\ }\textbf {\bibinfo {volume} {10}},\ \bibinfo {pages}
  {6477} (\bibinfo {year} {2014})}\BibitemShut {NoStop}%
\bibitem [{\citenamefont {Yan}\ and\ \citenamefont {Brady}(2015)}]{Yan2015}%
  \BibitemOpen
  \bibfield  {author} {\bibinfo {author} {\bibfnamefont {W.}~\bibnamefont
  {Yan}}\ and\ \bibinfo {author} {\bibfnamefont {J.~F.}\ \bibnamefont
  {Brady}},\ }\href {\doibase 10.1017/jfm.2015.621} {\bibfield  {journal}
  {\bibinfo  {journal} {Journal of Fluid Mechanics}\ }\textbf {\bibinfo
  {volume} {785}} (\bibinfo {year} {2015}),\ 10.1017/jfm.2015.621}\BibitemShut
  {NoStop}%
\bibitem [{\citenamefont {Ezhilan}\ \emph {et~al.}(2015)\citenamefont
  {Ezhilan}, \citenamefont {Alonso-Matilla},\ and\ \citenamefont
  {Saintillan}}]{Ezhilan2015}%
  \BibitemOpen
  \bibfield  {author} {\bibinfo {author} {\bibfnamefont {B.}~\bibnamefont
  {Ezhilan}}, \bibinfo {author} {\bibfnamefont {R.}~\bibnamefont
  {Alonso-Matilla}}, \ and\ \bibinfo {author} {\bibfnamefont {D.}~\bibnamefont
  {Saintillan}},\ }\href {\doibase 10.1017/jfm.2015.520} {\bibfield  {journal}
  {\bibinfo  {journal} {Journal of Fluid Mechanics}\ }\textbf {\bibinfo
  {volume} {781}} (\bibinfo {year} {2015}),\ 10.1017/jfm.2015.520}\BibitemShut
  {NoStop}%
\bibitem [{\citenamefont {Fily}\ \emph {et~al.}(2014)\citenamefont {Fily},
  \citenamefont {Baskaran},\ and\ \citenamefont {Hagan}}]{Fily2014}%
  \BibitemOpen
  \bibfield  {author} {\bibinfo {author} {\bibfnamefont {Y.}~\bibnamefont
  {Fily}}, \bibinfo {author} {\bibfnamefont {A.}~\bibnamefont {Baskaran}}, \
  and\ \bibinfo {author} {\bibfnamefont {M.~F.}\ \bibnamefont {Hagan}},\ }\href
  {\doibase 10.1039/C4SM00975D} {\bibfield  {journal} {\bibinfo  {journal}
  {Soft Matter}\ }\textbf {\bibinfo {volume} {10}},\ \bibinfo {pages} {5609}
  (\bibinfo {year} {2014})}\BibitemShut {NoStop}%
\bibitem [{\citenamefont {Winkler}\ \emph {et~al.}(2015)\citenamefont
  {Winkler}, \citenamefont {Wysocki},\ and\ \citenamefont
  {Gompper}}]{Winkler2015}%
  \BibitemOpen
  \bibfield  {author} {\bibinfo {author} {\bibfnamefont {R.~G.}\ \bibnamefont
  {Winkler}}, \bibinfo {author} {\bibfnamefont {A.}~\bibnamefont {Wysocki}}, \
  and\ \bibinfo {author} {\bibfnamefont {G.}~\bibnamefont {Gompper}},\ }\href
  {\doibase 10.1039/C5SM01412C} {\bibfield  {journal} {\bibinfo  {journal}
  {Soft Matter}\ }\textbf {\bibinfo {volume} {11}},\ \bibinfo {pages} {6680}
  (\bibinfo {year} {2015})}\BibitemShut {NoStop}%
\bibitem [{\citenamefont {Mallory}\ \emph
  {et~al.}(2014{\natexlab{a}})\citenamefont {Mallory}, \citenamefont {\ifmmode
  \check{S}\else \v{S}\fi{}ari\ifmmode~\acute{c}\else \'{c}\fi{}},
  \citenamefont {Valeriani},\ and\ \citenamefont {Cacciuto}}]{Mallory2014}%
  \BibitemOpen
  \bibfield  {author} {\bibinfo {author} {\bibfnamefont {S.~A.}\ \bibnamefont
  {Mallory}}, \bibinfo {author} {\bibfnamefont {A.}~\bibnamefont {\ifmmode
  \check{S}\else \v{S}\fi{}ari\ifmmode~\acute{c}\else \'{c}\fi{}}}, \bibinfo
  {author} {\bibfnamefont {C.}~\bibnamefont {Valeriani}}, \ and\ \bibinfo
  {author} {\bibfnamefont {A.}~\bibnamefont {Cacciuto}},\ }\href {\doibase
  10.1103/PhysRevE.89.052303} {\bibfield  {journal} {\bibinfo  {journal} {Phys.
  Rev. E}\ }\textbf {\bibinfo {volume} {89}},\ \bibinfo {pages} {052303}
  (\bibinfo {year} {2014}{\natexlab{a}})}\BibitemShut {NoStop}%
\bibitem [{\citenamefont {{Junot}}\ \emph {et~al.}(2017)\citenamefont
  {{Junot}}, \citenamefont {{Briand}}, \citenamefont {{Ledesma-Alonso}},\ and\
  \citenamefont {{Dauchot}}}]{Junot2017}%
  \BibitemOpen
  \bibfield  {author} {\bibinfo {author} {\bibfnamefont {G.}~\bibnamefont
  {{Junot}}}, \bibinfo {author} {\bibfnamefont {G.}~\bibnamefont {{Briand}}},
  \bibinfo {author} {\bibfnamefont {R.}~\bibnamefont {{Ledesma-Alonso}}}, \
  and\ \bibinfo {author} {\bibfnamefont {O.}~\bibnamefont {{Dauchot}}},\
  }\href@noop {} {\bibfield  {journal} {\bibinfo  {journal} {ArXiv e-prints}\ }
  (\bibinfo {year} {2017})},\ \Eprint {http://arxiv.org/abs/1703.04195}
  {arXiv:1703.04195 [cond-mat.soft]} \BibitemShut {NoStop}%
\bibitem [{\citenamefont {Reichhardt}\ and\ \citenamefont
  {Reichhardt}(2017)}]{Reichhardt2017}%
  \BibitemOpen
  \bibfield  {author} {\bibinfo {author} {\bibfnamefont {C.~O.}\ \bibnamefont
  {Reichhardt}}\ and\ \bibinfo {author} {\bibfnamefont {C.}~\bibnamefont
  {Reichhardt}},\ }\href {\doibase 10.1146/annurev-conmatphys-031016-025522}
  {\bibfield  {journal} {\bibinfo  {journal} {Annual Review of Condensed Matter
  Physics}\ }\textbf {\bibinfo {volume} {8}},\ \bibinfo {pages} {51} (\bibinfo
  {year} {2017})},\ \Eprint
  {http://arxiv.org/abs/https://doi.org/10.1146/annurev-conmatphys-031016-025522}
  {https://doi.org/10.1146/annurev-conmatphys-031016-025522} \BibitemShut
  {NoStop}%
\bibitem [{\citenamefont {Bechinger}\ \emph {et~al.}(2016)\citenamefont
  {Bechinger}, \citenamefont {Di~Leonardo}, \citenamefont {L\"owen},
  \citenamefont {Reichhardt}, \citenamefont {Volpe},\ and\ \citenamefont
  {Volpe}}]{Bechinger2016}%
  \BibitemOpen
  \bibfield  {author} {\bibinfo {author} {\bibfnamefont {C.}~\bibnamefont
  {Bechinger}}, \bibinfo {author} {\bibfnamefont {R.}~\bibnamefont
  {Di~Leonardo}}, \bibinfo {author} {\bibfnamefont {H.}~\bibnamefont
  {L\"owen}}, \bibinfo {author} {\bibfnamefont {C.}~\bibnamefont {Reichhardt}},
  \bibinfo {author} {\bibfnamefont {G.}~\bibnamefont {Volpe}}, \ and\ \bibinfo
  {author} {\bibfnamefont {G.}~\bibnamefont {Volpe}},\ }\href {\doibase
  10.1103/RevModPhys.88.045006} {\bibfield  {journal} {\bibinfo  {journal}
  {Rev. Mod. Phys.}\ }\textbf {\bibinfo {volume} {88}},\ \bibinfo {pages}
  {045006} (\bibinfo {year} {2016})}\BibitemShut {NoStop}%
\bibitem [{\citenamefont {Angelani}\ \emph {et~al.}(2009)\citenamefont
  {Angelani}, \citenamefont {Di~Leonardo},\ and\ \citenamefont
  {Ruocco}}]{Angelani2009}%
  \BibitemOpen
  \bibfield  {author} {\bibinfo {author} {\bibfnamefont {L.}~\bibnamefont
  {Angelani}}, \bibinfo {author} {\bibfnamefont {R.}~\bibnamefont
  {Di~Leonardo}}, \ and\ \bibinfo {author} {\bibfnamefont {G.}~\bibnamefont
  {Ruocco}},\ }\href {\doibase 10.1103/PhysRevLett.102.048104} {\bibfield
  {journal} {\bibinfo  {journal} {Phys. Rev. Lett.}\ }\textbf {\bibinfo
  {volume} {102}},\ \bibinfo {pages} {048104} (\bibinfo {year}
  {2009})}\BibitemShut {NoStop}%
\bibitem [{\citenamefont {Angelani}\ \emph {et~al.}(2011)\citenamefont
  {Angelani}, \citenamefont {Costanzo},\ and\ \citenamefont
  {Leonardo}}]{Angelani2011}%
  \BibitemOpen
  \bibfield  {author} {\bibinfo {author} {\bibfnamefont {L.}~\bibnamefont
  {Angelani}}, \bibinfo {author} {\bibfnamefont {A.}~\bibnamefont {Costanzo}},
  \ and\ \bibinfo {author} {\bibfnamefont {R.~D.}\ \bibnamefont {Leonardo}},\
  }\href {http://stacks.iop.org/0295-5075/96/i=6/a=68002} {\bibfield  {journal}
  {\bibinfo  {journal} {EPL (Europhysics Letters)}\ }\textbf {\bibinfo {volume}
  {96}},\ \bibinfo {pages} {68002} (\bibinfo {year} {2011})}\BibitemShut
  {NoStop}%
\bibitem [{\citenamefont {Di~Leonardo}\ \emph {et~al.}(2010)\citenamefont
  {Di~Leonardo}, \citenamefont {Angelani}, \citenamefont {Dell’Arciprete},
  \citenamefont {Ruocco}, \citenamefont {Iebba}, \citenamefont {Schippa},
  \citenamefont {Conte}, \citenamefont {Mecarini}, \citenamefont {De~Angelis},\
  and\ \citenamefont {Di~Fabrizio}}]{DiLeonardo2010}%
  \BibitemOpen
  \bibfield  {author} {\bibinfo {author} {\bibfnamefont {R.}~\bibnamefont
  {Di~Leonardo}}, \bibinfo {author} {\bibfnamefont {L.}~\bibnamefont
  {Angelani}}, \bibinfo {author} {\bibfnamefont {D.}~\bibnamefont
  {Dell’Arciprete}}, \bibinfo {author} {\bibfnamefont {G.}~\bibnamefont
  {Ruocco}}, \bibinfo {author} {\bibfnamefont {V.}~\bibnamefont {Iebba}},
  \bibinfo {author} {\bibfnamefont {S.}~\bibnamefont {Schippa}}, \bibinfo
  {author} {\bibfnamefont {M.~P.}\ \bibnamefont {Conte}}, \bibinfo {author}
  {\bibfnamefont {F.}~\bibnamefont {Mecarini}}, \bibinfo {author}
  {\bibfnamefont {F.}~\bibnamefont {De~Angelis}}, \ and\ \bibinfo {author}
  {\bibfnamefont {E.}~\bibnamefont {Di~Fabrizio}},\ }\href {\doibase
  10.1073/pnas.0910426107} {\ \textbf {\bibinfo {volume} {107}},\ \bibinfo
  {pages} {9541} (\bibinfo {year} {2010})},\ \Eprint
  {http://arxiv.org/abs/http://www.pnas.org/content/107/21/9541.full.pdf}
  {http://www.pnas.org/content/107/21/9541.full.pdf} \BibitemShut {NoStop}%
\bibitem [{\citenamefont {Galajda}\ \emph {et~al.}(2007)\citenamefont
  {Galajda}, \citenamefont {Keymer}, \citenamefont {Chaikin},\ and\
  \citenamefont {Austin}}]{Galajda2007}%
  \BibitemOpen
  \bibfield  {author} {\bibinfo {author} {\bibfnamefont {P.}~\bibnamefont
  {Galajda}}, \bibinfo {author} {\bibfnamefont {J.}~\bibnamefont {Keymer}},
  \bibinfo {author} {\bibfnamefont {P.}~\bibnamefont {Chaikin}}, \ and\
  \bibinfo {author} {\bibfnamefont {R.}~\bibnamefont {Austin}},\ }\href
  {\doibase 10.1128/JB.01033-07} {\bibfield  {journal} {\bibinfo  {journal}
  {Journal of Bacteriology}\ }\textbf {\bibinfo {volume} {189}},\ \bibinfo
  {pages} {8704} (\bibinfo {year} {2007})},\ \Eprint
  {http://arxiv.org/abs/http://jb.asm.org/content/189/23/8704.full.pdf+html}
  {http://jb.asm.org/content/189/23/8704.full.pdf+html} \BibitemShut {NoStop}%
\bibitem [{\citenamefont {Kaiser}\ \emph {et~al.}(2014)\citenamefont {Kaiser},
  \citenamefont {Peshkov}, \citenamefont {Sokolov}, \citenamefont {ten Hagen},
  \citenamefont {L\"owen},\ and\ \citenamefont {Aranson}}]{Kaiser2014}%
  \BibitemOpen
  \bibfield  {author} {\bibinfo {author} {\bibfnamefont {A.}~\bibnamefont
  {Kaiser}}, \bibinfo {author} {\bibfnamefont {A.}~\bibnamefont {Peshkov}},
  \bibinfo {author} {\bibfnamefont {A.}~\bibnamefont {Sokolov}}, \bibinfo
  {author} {\bibfnamefont {B.}~\bibnamefont {ten Hagen}}, \bibinfo {author}
  {\bibfnamefont {H.}~\bibnamefont {L\"owen}}, \ and\ \bibinfo {author}
  {\bibfnamefont {I.~S.}\ \bibnamefont {Aranson}},\ }\href {\doibase
  10.1103/PhysRevLett.112.158101} {\bibfield  {journal} {\bibinfo  {journal}
  {Phys. Rev. Lett.}\ }\textbf {\bibinfo {volume} {112}},\ \bibinfo {pages}
  {158101} (\bibinfo {year} {2014})}\BibitemShut {NoStop}%
\bibitem [{\citenamefont {Wan}\ \emph {et~al.}(2008)\citenamefont {Wan},
  \citenamefont {Olson~Reichhardt}, \citenamefont {Nussinov},\ and\
  \citenamefont {Reichhardt}}]{Wan2008}%
  \BibitemOpen
  \bibfield  {author} {\bibinfo {author} {\bibfnamefont {M.~B.}\ \bibnamefont
  {Wan}}, \bibinfo {author} {\bibfnamefont {C.~J.}\ \bibnamefont
  {Olson~Reichhardt}}, \bibinfo {author} {\bibfnamefont {Z.}~\bibnamefont
  {Nussinov}}, \ and\ \bibinfo {author} {\bibfnamefont {C.}~\bibnamefont
  {Reichhardt}},\ }\href {\doibase 10.1103/PhysRevLett.101.018102} {\bibfield
  {journal} {\bibinfo  {journal} {Phys. Rev. Lett.}\ }\textbf {\bibinfo
  {volume} {101}},\ \bibinfo {pages} {018102} (\bibinfo {year}
  {2008})}\BibitemShut {NoStop}%
\bibitem [{\citenamefont {Tailleur}\ and\ \citenamefont
  {Cates}(2009)}]{Tailleur2009}%
  \BibitemOpen
  \bibfield  {author} {\bibinfo {author} {\bibfnamefont {J.}~\bibnamefont
  {Tailleur}}\ and\ \bibinfo {author} {\bibfnamefont {M.~E.}\ \bibnamefont
  {Cates}},\ }\href {\doibase 10.1209/0295-5075/86/60002} {\bibfield  {journal}
  {\bibinfo  {journal} {EPL (Europhysics Letters)}\ }\textbf {\bibinfo {volume}
  {86}},\ \bibinfo {pages} {60002} (\bibinfo {year} {2009})}\BibitemShut
  {NoStop}%
\bibitem [{\citenamefont {Magnasco}(1993)}]{Magnasco1993}%
  \BibitemOpen
  \bibfield  {author} {\bibinfo {author} {\bibfnamefont {M.~O.}\ \bibnamefont
  {Magnasco}},\ }\href {\doibase 10.1103/PhysRevLett.71.1477} {\bibfield
  {journal} {\bibinfo  {journal} {Phys. Rev. Lett.}\ }\textbf {\bibinfo
  {volume} {71}},\ \bibinfo {pages} {1477} (\bibinfo {year}
  {1993})}\BibitemShut {NoStop}%
\bibitem [{\citenamefont {J\"ulicher}\ \emph {et~al.}(1997)\citenamefont
  {J\"ulicher}, \citenamefont {Ajdari},\ and\ \citenamefont
  {Prost}}]{Julicher1997}%
  \BibitemOpen
  \bibfield  {author} {\bibinfo {author} {\bibfnamefont {F.}~\bibnamefont
  {J\"ulicher}}, \bibinfo {author} {\bibfnamefont {A.}~\bibnamefont {Ajdari}},
  \ and\ \bibinfo {author} {\bibfnamefont {J.}~\bibnamefont {Prost}},\ }\href
  {\doibase 10.1103/RevModPhys.69.1269} {\bibfield  {journal} {\bibinfo
  {journal} {Rev. Mod. Phys.}\ }\textbf {\bibinfo {volume} {69}},\ \bibinfo
  {pages} {1269} (\bibinfo {year} {1997})}\BibitemShut {NoStop}%
\bibitem [{\citenamefont {Feynman}\ \emph {et~al.}(1966)\citenamefont
  {Feynman}, \citenamefont {Leighton},\ and\ \citenamefont
  {Sands}}]{Feynman1966}%
  \BibitemOpen
  \bibfield  {author} {\bibinfo {author} {\bibfnamefont {R.~P.}\ \bibnamefont
  {Feynman}}, \bibinfo {author} {\bibfnamefont {R.~B.}\ \bibnamefont
  {Leighton}}, \ and\ \bibinfo {author} {\bibfnamefont {M.}~\bibnamefont
  {Sands}},\ }\href@noop {} {\emph {\bibinfo {title} {The Feynman Lectures on
  Physics, Vol. I}}}\ (\bibinfo  {publisher} {Addison-Wesley},\ \bibinfo
  {address} {Reading, Mass},\ \bibinfo {year} {1966})\BibitemShut {NoStop}%
\bibitem [{\citenamefont {Marchetti}\ \emph {et~al.}(2013)\citenamefont
  {Marchetti}, \citenamefont {Joanny}, \citenamefont {Ramaswamy}, \citenamefont
  {Liverpool}, \citenamefont {Prost}, \citenamefont {Rao},\ and\ \citenamefont
  {Simha}}]{Marchetti2013rev}%
  \BibitemOpen
  \bibfield  {author} {\bibinfo {author} {\bibfnamefont {M.~C.}\ \bibnamefont
  {Marchetti}}, \bibinfo {author} {\bibfnamefont {J.~F.}\ \bibnamefont
  {Joanny}}, \bibinfo {author} {\bibfnamefont {S.}~\bibnamefont {Ramaswamy}},
  \bibinfo {author} {\bibfnamefont {T.~B.}\ \bibnamefont {Liverpool}}, \bibinfo
  {author} {\bibfnamefont {J.}~\bibnamefont {Prost}}, \bibinfo {author}
  {\bibfnamefont {M.}~\bibnamefont {Rao}}, \ and\ \bibinfo {author}
  {\bibfnamefont {R.~A.}\ \bibnamefont {Simha}},\ }\href {\doibase
  10.1103/RevModPhys.85.1143} {\bibfield  {journal} {\bibinfo  {journal} {Rev.
  Mod. Phys.}\ }\textbf {\bibinfo {volume} {85}},\ \bibinfo {pages} {1143}
  (\bibinfo {year} {2013})}\BibitemShut {NoStop}%
\bibitem [{\citenamefont {Almonacid}\ \emph {et~al.}(2015)\citenamefont
  {Almonacid}, \citenamefont {Ahmed}, \citenamefont {Bussonnier}, \citenamefont
  {Mailly}, \citenamefont {Betz}, \citenamefont {Voituriez}, \citenamefont
  {Gov},\ and\ \citenamefont {Verlhac}}]{Almonacid2015}%
  \BibitemOpen
  \bibfield  {author} {\bibinfo {author} {\bibfnamefont {M.}~\bibnamefont
  {Almonacid}}, \bibinfo {author} {\bibfnamefont {W.~W.}\ \bibnamefont
  {Ahmed}}, \bibinfo {author} {\bibfnamefont {M.}~\bibnamefont {Bussonnier}},
  \bibinfo {author} {\bibfnamefont {P.}~\bibnamefont {Mailly}}, \bibinfo
  {author} {\bibfnamefont {T.}~\bibnamefont {Betz}}, \bibinfo {author}
  {\bibfnamefont {R.}~\bibnamefont {Voituriez}}, \bibinfo {author}
  {\bibfnamefont {N.~S.}\ \bibnamefont {Gov}}, \ and\ \bibinfo {author}
  {\bibfnamefont {M.-H.}\ \bibnamefont {Verlhac}},\ }\href {\doibase
  10.1038/ncb3131} {\bibfield  {journal} {\bibinfo  {journal} {Nature Cell
  Biology}\ }\textbf {\bibinfo {volume} {17}},\ \bibinfo {pages} {470}
  (\bibinfo {year} {2015})}\BibitemShut {NoStop}%
\bibitem [{Note1()}]{Note1}%
  \BibitemOpen
  \bibinfo {note} {The vesicles are driven by molecular motors that walk on
  actin filaments. The actin network that fills the volume of the oocyte is
  dynamic, allowing the viscous motion of the nucleus on long time-scales. The
  momentum that is imparted by the active vesicles to the nucleus could be
  balanced by the forces that are transmitted through the actin network to the
  rigid cortex of the oocyte, and to the rest of the lab.}\BibitemShut {Stop}%
\bibitem [{\citenamefont {Lozano}\ \emph {et~al.}(2016)\citenamefont {Lozano},
  \citenamefont {ten Hagen}, \citenamefont {L{\"o}wen},\ and\ \citenamefont
  {Bechinger}}]{Lozano2016}%
  \BibitemOpen
  \bibfield  {author} {\bibinfo {author} {\bibfnamefont {C.}~\bibnamefont
  {Lozano}}, \bibinfo {author} {\bibfnamefont {B.}~\bibnamefont {ten Hagen}},
  \bibinfo {author} {\bibfnamefont {H.}~\bibnamefont {L{\"o}wen}}, \ and\
  \bibinfo {author} {\bibfnamefont {C.}~\bibnamefont {Bechinger}},\ }\href
  {http://dx.doi.org/10.1038/ncomms12828} {\bibfield  {journal} {\bibinfo
  {journal} {Nature Communications}\ }\textbf {\bibinfo {volume} {7}},\
  \bibinfo {pages} {12828 EP } (\bibinfo {year} {2016})},\ \bibinfo {note}
  {article}\BibitemShut {NoStop}%
\bibitem [{\citenamefont {Palacci}\ \emph {et~al.}(2013)\citenamefont
  {Palacci}, \citenamefont {Sacanna}, \citenamefont {Steinberg}, \citenamefont
  {Pine},\ and\ \citenamefont {Chaikin}}]{Palacci2013}%
  \BibitemOpen
  \bibfield  {author} {\bibinfo {author} {\bibfnamefont {J.}~\bibnamefont
  {Palacci}}, \bibinfo {author} {\bibfnamefont {S.}~\bibnamefont {Sacanna}},
  \bibinfo {author} {\bibfnamefont {A.~P.}\ \bibnamefont {Steinberg}}, \bibinfo
  {author} {\bibfnamefont {D.~J.}\ \bibnamefont {Pine}}, \ and\ \bibinfo
  {author} {\bibfnamefont {P.~M.}\ \bibnamefont {Chaikin}},\ }\href {\doibase
  10.1126/science.1230020} {\bibfield  {journal} {\bibinfo  {journal}
  {Science}\ }\textbf {\bibinfo {volume} {339}},\ \bibinfo {pages} {936}
  (\bibinfo {year} {2013})},\ \Eprint
  {http://arxiv.org/abs/http://science.sciencemag.org/content/339/6122/936.full.pdf}
  {http://science.sciencemag.org/content/339/6122/936.full.pdf} \BibitemShut
  {NoStop}%
\bibitem [{\citenamefont {Buttinoni}\ \emph {et~al.}(2013)\citenamefont
  {Buttinoni}, \citenamefont {Bialk\'e}, \citenamefont {K\"ummel},
  \citenamefont {L\"owen}, \citenamefont {Bechinger},\ and\ \citenamefont
  {Speck}}]{Buttinoni2013}%
  \BibitemOpen
  \bibfield  {author} {\bibinfo {author} {\bibfnamefont {I.}~\bibnamefont
  {Buttinoni}}, \bibinfo {author} {\bibfnamefont {J.}~\bibnamefont {Bialk\'e}},
  \bibinfo {author} {\bibfnamefont {F.}~\bibnamefont {K\"ummel}}, \bibinfo
  {author} {\bibfnamefont {H.}~\bibnamefont {L\"owen}}, \bibinfo {author}
  {\bibfnamefont {C.}~\bibnamefont {Bechinger}}, \ and\ \bibinfo {author}
  {\bibfnamefont {T.}~\bibnamefont {Speck}},\ }\href {\doibase
  10.1103/PhysRevLett.110.238301} {\bibfield  {journal} {\bibinfo  {journal}
  {Phys. Rev. Lett.}\ }\textbf {\bibinfo {volume} {110}},\ \bibinfo {pages}
  {238301} (\bibinfo {year} {2013})}\BibitemShut {NoStop}%
\bibitem [{\citenamefont {Schnitzer}(1993)}]{Schnitzer1993}%
  \BibitemOpen
  \bibfield  {author} {\bibinfo {author} {\bibfnamefont {M.~J.}\ \bibnamefont
  {Schnitzer}},\ }\href {\doibase 10.1103/PhysRevE.48.2553} {\bibfield
  {journal} {\bibinfo  {journal} {Physical Review E}\ }\textbf {\bibinfo
  {volume} {48}},\ \bibinfo {pages} {2553} (\bibinfo {year}
  {1993})}\BibitemShut {NoStop}%
\bibitem [{\citenamefont {Lee}(2013)}]{Lee2013}%
  \BibitemOpen
  \bibfield  {author} {\bibinfo {author} {\bibfnamefont {C.~F.}\ \bibnamefont
  {Lee}},\ }\href {http://stacks.iop.org/1367-2630/15/i=5/a=055007} {\bibfield
  {journal} {\bibinfo  {journal} {New Journal of Physics}\ }\textbf {\bibinfo
  {volume} {15}},\ \bibinfo {pages} {055007} (\bibinfo {year}
  {2013})}\BibitemShut {NoStop}%
\bibitem [{\citenamefont {Elgeti}\ and\ \citenamefont
  {Gompper}(2015)}]{Elgeti2015}%
  \BibitemOpen
  \bibfield  {author} {\bibinfo {author} {\bibfnamefont {J.}~\bibnamefont
  {Elgeti}}\ and\ \bibinfo {author} {\bibfnamefont {G.}~\bibnamefont
  {Gompper}},\ }\href {http://stacks.iop.org/0295-5075/109/i=5/a=58003}
  {\bibfield  {journal} {\bibinfo  {journal} {EPL (Europhysics Letters)}\
  }\textbf {\bibinfo {volume} {109}},\ \bibinfo {pages} {58003} (\bibinfo
  {year} {2015})}\BibitemShut {NoStop}%
\bibitem [{\citenamefont {Tailleur}\ and\ \citenamefont
  {Cates}(2008)}]{Tailleur2008}%
  \BibitemOpen
  \bibfield  {author} {\bibinfo {author} {\bibfnamefont {J.}~\bibnamefont
  {Tailleur}}\ and\ \bibinfo {author} {\bibfnamefont {M.~E.}\ \bibnamefont
  {Cates}},\ }\href {\doibase 10.1103/PhysRevLett.100.218103} {\bibfield
  {journal} {\bibinfo  {journal} {Physical review letters}\ }\textbf {\bibinfo
  {volume} {100}},\ \bibinfo {pages} {218103} (\bibinfo {year} {2008})},\
  \Eprint {http://arxiv.org/abs/0803.1069} {arXiv:0803.1069} \BibitemShut
  {NoStop}%
\bibitem [{com()}]{comment1}%
  \BibitemOpen
  \href@noop {} {}\bibinfo {note} {While the bulk equations are identical to
  the ones used by Schnitzer et. al. (1993) to study the flux in open systems,
  here we study the steady state force on a piston by finding the density near
  boundaries, which is described by additional equations due to the edge
  discontinuity of the density.}\BibitemShut {Stop}%
\bibitem [{\citenamefont {Cates}\ and\ \citenamefont
  {Tailleur}(2015)}]{Cates2015}%
  \BibitemOpen
  \bibfield  {author} {\bibinfo {author} {\bibfnamefont {M.~E.}\ \bibnamefont
  {Cates}}\ and\ \bibinfo {author} {\bibfnamefont {J.}~\bibnamefont
  {Tailleur}},\ }\href {\doibase 10.1146/annurev-conmatphys-031214-014710}
  {\bibfield  {journal} {\bibinfo  {journal} {Annual Review of Condensed Matter
  Physics}\ }\textbf {\bibinfo {volume} {6}},\ \bibinfo {pages} {219} (\bibinfo
  {year} {2015})},\ \Eprint
  {http://arxiv.org/abs/http://dx.doi.org/10.1146/annurev-conmatphys-031214-014710}
  {http://dx.doi.org/10.1146/annurev-conmatphys-031214-014710} \BibitemShut
  {NoStop}%
\bibitem [{\citenamefont {Elgeti}\ and\ \citenamefont
  {Gompper}(2013)}]{Elgeti2013}%
  \BibitemOpen
  \bibfield  {author} {\bibinfo {author} {\bibfnamefont {J.}~\bibnamefont
  {Elgeti}}\ and\ \bibinfo {author} {\bibfnamefont {G.}~\bibnamefont
  {Gompper}},\ }\href {http://stacks.iop.org/0295-5075/101/i=4/a=48003}
  {\bibfield  {journal} {\bibinfo  {journal} {EPL (Europhysics Letters)}\
  }\textbf {\bibinfo {volume} {101}},\ \bibinfo {pages} {48003} (\bibinfo
  {year} {2013})}\BibitemShut {NoStop}%
\bibitem [{\citenamefont {Fily}\ \emph {et~al.}(2015)\citenamefont {Fily},
  \citenamefont {Baskaran},\ and\ \citenamefont {Hagan}}]{Fily2015}%
  \BibitemOpen
  \bibfield  {author} {\bibinfo {author} {\bibfnamefont {Y.}~\bibnamefont
  {Fily}}, \bibinfo {author} {\bibfnamefont {A.}~\bibnamefont {Baskaran}}, \
  and\ \bibinfo {author} {\bibfnamefont {M.~F.}\ \bibnamefont {Hagan}},\ }\href
  {\doibase 10.1103/PhysRevE.91.012125} {\bibfield  {journal} {\bibinfo
  {journal} {Phys. Rev. E}\ }\textbf {\bibinfo {volume} {91}},\ \bibinfo
  {pages} {012125} (\bibinfo {year} {2015})}\BibitemShut {NoStop}%
\bibitem [{\citenamefont {Harder}\ \emph {et~al.}(2014)\citenamefont {Harder},
  \citenamefont {Mallory}, \citenamefont {Tung}, \citenamefont {Valeriani},\
  and\ \citenamefont {Cacciuto}}]{Harder2014}%
  \BibitemOpen
  \bibfield  {author} {\bibinfo {author} {\bibfnamefont {J.}~\bibnamefont
  {Harder}}, \bibinfo {author} {\bibfnamefont {S.~A.}\ \bibnamefont {Mallory}},
  \bibinfo {author} {\bibfnamefont {C.}~\bibnamefont {Tung}}, \bibinfo {author}
  {\bibfnamefont {C.}~\bibnamefont {Valeriani}}, \ and\ \bibinfo {author}
  {\bibfnamefont {A.}~\bibnamefont {Cacciuto}},\ }\href {\doibase
  10.1063/1.4900720} {\bibfield  {journal} {\bibinfo  {journal} {The Journal of
  Chemical Physics}\ }\textbf {\bibinfo {volume} {141}},\ \bibinfo {pages}
  {194901} (\bibinfo {year} {2014})},\ \Eprint
  {http://arxiv.org/abs/http://dx.doi.org/10.1063/1.4900720}
  {http://dx.doi.org/10.1063/1.4900720} \BibitemShut {NoStop}%
\bibitem [{\citenamefont {Mallory}\ \emph
  {et~al.}(2014{\natexlab{b}})\citenamefont {Mallory}, \citenamefont
  {Valeriani},\ and\ \citenamefont {Cacciuto}}]{Mallory2014b}%
  \BibitemOpen
  \bibfield  {author} {\bibinfo {author} {\bibfnamefont {S.~A.}\ \bibnamefont
  {Mallory}}, \bibinfo {author} {\bibfnamefont {C.}~\bibnamefont {Valeriani}},
  \ and\ \bibinfo {author} {\bibfnamefont {A.}~\bibnamefont {Cacciuto}},\
  }\href {\doibase 10.1103/PhysRevE.90.032309} {\bibfield  {journal} {\bibinfo
  {journal} {Phys. Rev. E}\ }\textbf {\bibinfo {volume} {90}},\ \bibinfo
  {pages} {032309} (\bibinfo {year} {2014}{\natexlab{b}})}\BibitemShut
  {NoStop}%
\bibitem [{\citenamefont {Smallenburg}\ and\ \citenamefont
  {L\"owen}(2015)}]{Smallenburg2015}%
  \BibitemOpen
  \bibfield  {author} {\bibinfo {author} {\bibfnamefont {F.}~\bibnamefont
  {Smallenburg}}\ and\ \bibinfo {author} {\bibfnamefont {H.}~\bibnamefont
  {L\"owen}},\ }\href {\doibase 10.1103/PhysRevE.92.032304} {\bibfield
  {journal} {\bibinfo  {journal} {Phys. Rev. E}\ }\textbf {\bibinfo {volume}
  {92}},\ \bibinfo {pages} {032304} (\bibinfo {year} {2015})}\BibitemShut
  {NoStop}%
\bibitem [{\citenamefont {Cates}\ and\ \citenamefont
  {Tailleur}(2013)}]{Cates2013ABPvsRTP}%
  \BibitemOpen
  \bibfield  {author} {\bibinfo {author} {\bibfnamefont {M.~E.}\ \bibnamefont
  {Cates}}\ and\ \bibinfo {author} {\bibfnamefont {J.}~\bibnamefont
  {Tailleur}},\ }\href {http://stacks.iop.org/0295-5075/101/i=2/a=20010}
  {\bibfield  {journal} {\bibinfo  {journal} {EPL (Europhysics Letters)}\
  }\textbf {\bibinfo {volume} {101}},\ \bibinfo {pages} {20010} (\bibinfo
  {year} {2013})}\BibitemShut {NoStop}%
\bibitem [{\citenamefont {Solon}\ \emph
  {et~al.}(2015{\natexlab{c}})\citenamefont {Solon}, \citenamefont {Cates},\
  and\ \citenamefont {Tailleur}}]{Solon2015ABPRTP}%
  \BibitemOpen
  \bibfield  {author} {\bibinfo {author} {\bibfnamefont {A.~P.}\ \bibnamefont
  {Solon}}, \bibinfo {author} {\bibfnamefont {M.~E.}\ \bibnamefont {Cates}}, \
  and\ \bibinfo {author} {\bibfnamefont {J.}~\bibnamefont {Tailleur}},\ }\href
  {\doibase 10.1140/epjst/e2015-02457-0} {\bibfield  {journal} {\bibinfo
  {journal} {The European Physical Journal Special Topics}\ }\textbf {\bibinfo
  {volume} {224}},\ \bibinfo {pages} {1231} (\bibinfo {year}
  {2015}{\natexlab{c}})}\BibitemShut {NoStop}%
\end{thebibliography}
%

\end{document}